\documentclass[12pt]{article}
\usepackage[T1]{fontenc}
\usepackage[utf8]{inputenc}  
\usepackage{amsmath,amssymb,amsthm}
\usepackage{booktabs}
\usepackage{caption}
\usepackage{float}
\usepackage{multirow}
\usepackage{subfig}
\usepackage{tabularx}
\usepackage{multicol}
\usepackage[svgnames,dvipsnames,table]{xcolor}
\usepackage{pgfplots}
\pgfplotsset{compat=newest}
\usetikzlibrary{spy,backgrounds}
\usepackage{pgfplotstable}
\usepackage{enumerate}

\def\inh{\vskip 0.075truein \noindent\hangindent=12 pt \hangafter=1}

\theoremstyle{remark}

\newcommand{\bte}{\begin{quote}\begin{theorem}}
		\newcommand{\ete}[1]{\label{#1}\end{theorem}\end{quote}}
\newcommand{\bcom}{\begin{quote}\end{quote}}
\newcommand{\bex}{\begin{quote}\begin{example}}
		\newcommand{\eex}[1]{\label{#1}\end{example}\end{quote}}
\newcommand{\bcon}{\begin{quote}\begin{conclusion}}
		\newcommand{\econ}[1]{\label{#1}\end{conclusion}\end{quote}}
\newcommand{\bdefi}{\begin{quote}\begin{definition}}
		\newcommand{\edefi}[1]{\label{#1}\end{definition}\end{quote}}

\newcommand{\blem}{\begin{quote}\begin{lemma}}
		\newcommand{\elem}[1]{\label{#1}\end{lemma}\end{quote}}

\newcommand{\bpr}{\begin{quote}\begin{problem}}
		\newcommand{\epr}[1]{\label{#1}\end{problem}\end{quote}}

\newcommand{\f}{\frac}

\newcommand{\beq}{\begin{eqnarray}}
\newcommand{\eeq}[1]{\label{#1}\end{eqnarray}}

\newcommand{\bfi}{\begin{figure}[24]}
	\newcommand{\efi}[1]{\caption{\label{#1}}\end{figure}}

\label{key}

\newcommand{\CM}{{\cal M}}

\newcommand{\CP}{{\cal P}}

\newcommand{\CV}{{\cal V}}
\newcommand{\CW}{{\cal W}}

\newcommand\D{\,\mathrm{d}}
\newcommand\I{\mathrm{i}}

\newcommand{\bexe}{\begin{quote}\begin{exercise}\inh}
		\newcommand{\eexe}[1]{\label{#1}\end{exercise}\end{quote}}

\title{Alternating strain regimes for failure propagation in flexural systems}
\author{M. Garau\footnote{Keele University, School of Computing and Mathematics, Keele, ST5 5BG, UK}, M.J. Nieves$^{*,}$\footnote{University of Cagliari, Department of Mechanical, Chemical and Material Engineering, Cagliari, 09123, Italy} and I.S. Jones\footnote{Mechanical Engineering and Materials Research Centre, Liverpool John Moores University, James Parsons Building, Byrom Street, Liverpool L3 3AF, U.K.}}
\date{}

\begin{document}
	\maketitle



\begin{abstract}
We consider both analytical and numerical studies of a  steady-state fracture process inside a discrete mass-beam structure, composed of periodically placed masses connected by Euler-Bernoulli beams.
A fault inside the structure is assumed to propagate with a constant speed and this occurs as a result of the action of a remote sinusoidal, mechanical load. The established regime of fracture corresponds to the case of an alternating {generalised} strain regime.
The model is reduced to 
a Wiener-Hopf equation and its solution is presented. 
We determine the minimum feeding wave energy required for the steady-state fracture process to occur. In addition, we identify the dynamic features of the structure during the steady-state fracture regime.
A transient analysis of this problem is also presented, where the existence of steady-state fracture regimes, revealed by the analytical model, are verified and the associated transient features of this process are discussed.
\end{abstract}

{\bf Keywords\rm}: Discrete periodic media, mass-beam structures, fracture, Wiener-Hopf technique, numerical simulations.
\section{Introduction}

\begin{figure}[tp]
\centering
\includegraphics[width=14cm]{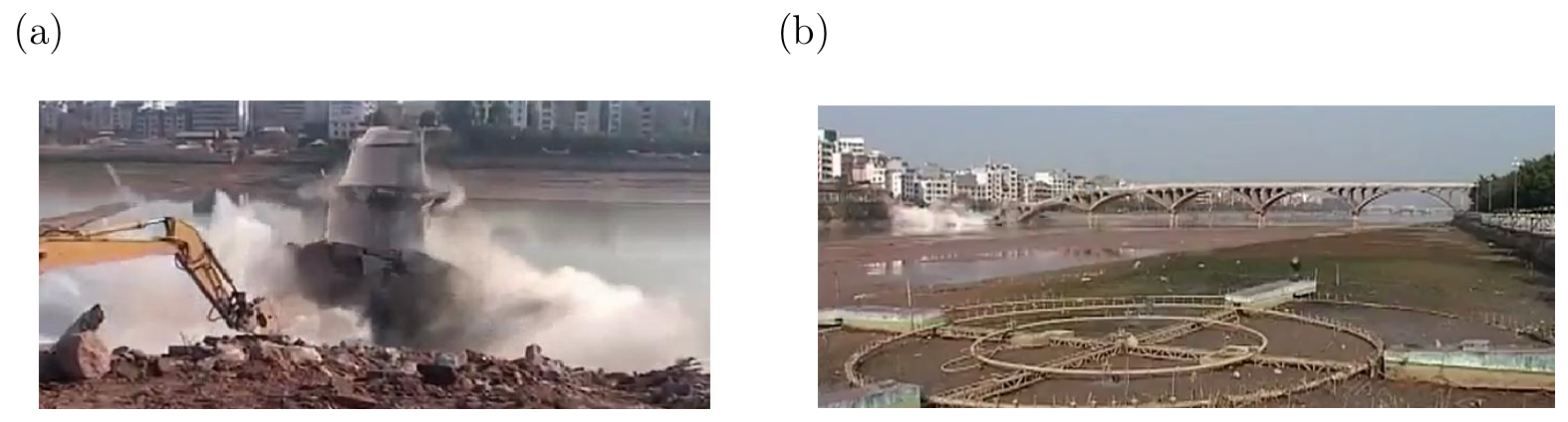}
\caption[]{(a) The demolition of a bridge in Guangxi, China.  The excavator in this picture was used to initiate a failure at one end of the bridge. The figure shows the resulting periodic failure process propagating through the structure, where the supporting columns fail sequentially. 
An alternative view of this process is presented in (b). (online version in colour)
}
\label{Picture}
\end{figure}

The modelling of periodic flexural materials is a useful tool in understanding the behaviour of structures commonly found in civil engineering, such as buildings, bridges, rooftops, pipeline systems and many more. The need to understand the response of these structures, as shown in Figure~\ref{Picture},  is greater when failure initiates and propagates through the system. 

We present a simplified analytical and numerical model to represent the failure of a long bridge or a rooftop. The structure is modelled by Euler-Bernoulli beams connecting periodically placed masses, where the failure is assumed to propagate steadily within the structure.

The modelling of failure in structured materials has been widely developed for the analysis of mass-spring systems. These models help in describing several microstructural processes which occur in the fracture of materials \cite{MG}.
Dynamic Mode III fracture of a square cell lattice, composed of massless springs connecting periodically placed masses, has been considered in \cite{LS1}.  For discrete mass-spring systems which undergo phase transition processes at a uniform rate, see \cite{LS2}.  Dynamic fracture modes in elastic triangular lattices were treated in \cite{LS3} for a homogeneous lattice and in \cite{NMJM} for an anisotropic lattice. 
In these models, the micro-level processes involved in the fracture phenomenon 
can be identified, including effects such as wave radiation occurring when failure propagates.
The introduction of inertial links in a lattice brings new features in terms of wave radiation processes that accompany the crack growth, as demonstrated for Mode III fracture inside an inertial square cell lattice in \cite{LS4}.

In addition to the analytical models considered in \cite{LS1, LS2, LS3, LS4}, numerical modelling of Mode I  and II crack growth inside strips of triangular lattice has been carried out in \cite{SA}, where some surprising periodic patterns of fault propagation were observed.

The analytical approach developed in \cite{LSbook} for treating phase transition in a structured material utilises the Fourier transform of the governing equations with respect to a moving coordinate that follows the position of the phase transition front. The problem can then be reduced to a Wiener-Hopf equation along the line containing the defect, where the conditions ahead of and behind the phase transition front differ.
This functional equation contains information about the dynamics of the medium on both macro- and micro-scales. The application of the Wiener-Hopf technique to solving dynamic problems for lattices with defects is of wide utility and has also been employed in  \cite{Sharma1, Sharma2, Sharma3, Sharma4, Sharma5} in  analysing scattering and diffraction of waves by defects in several types of periodic media.

Lattices, composed of fundamental mechanical elements, can easily be designed to produce a variety of media capable of controlling the flow of waves for different applications. In particular, the introduction of structural heterogeneities may influence the admissible regimes where steady crack propagation is possible. The effects of inhomogeneities on Mode III crack propagation in 
two-dimensional lattices
have been analysed in \cite{MMS}. The dynamic  failure of dissimilar chains  has been analysed in \cite{Berinskii, Gorbushinetaldissimilar}.

Periodically distributed  inhomogeneities in mass-spring lattices have been shown to enhance fracture propagation 
in \cite{MMS1}. There, a structured interface was shown to induce a  failure mechanism caused by a ``knife wave'' or localised deformation surrounding the crack, capable of sustaining  crack propagation within the interface. 
Numerical simulations of a crack growing through a high-contrast interface were later carried out in \cite{SMM}.
Moreover, these numerical simulations revealed non-steady failure regimes such as clustering, where a crack may propagate in an unusual way.

Other failure processes propagating in a lattice have been treated in \cite{MMS2}, where the extraction of a mass-spring chain due to a point force was considered.
The study  of a propagating bridge crack in an inhomogeneous lattice has been carried out in \cite{MMS3}. 
Different fracture criteria, independent of or dependent on time, can also produce a variety of fracture patterns within lattice systems and can  affect the existence  of admissible steady-state fracture regimes.  The impact of time dependent fracture criteria on the failure of   one-dimensional mass-spring chains has been investigated in \cite{GGM}. 
  
The important question of admissibility of unstable and stable crack regimes for cracks propagating in discrete  periodic media has been discussed in \cite{MG, LSbook, M}.
The existence of admissible low velocity steady-state failure regimes in a mass-spring square cell lattice \cite{LSbook}  has recently been re-addressed in \cite{GM}.  The admissibility of crack propagation regimes can also be investigated for different loading conditions.
The influence of a moving load on Mode III fracture regimes in a discrete mass-spring structure  has been analytically and numerically studied in \cite{GM2}. A transient analysis of thermal shock induced  crack propagation in triangular lattices has also been carried out in \cite{Cartaetal, Trevisan} and in \cite{Tallerico}, where cracks with chiral coatings were considered.

Failure regimes in discrete structures with non-local interactions and their effects on the wave radiation processes are presented in \cite{GM3} and for media with additional non-linearities in the connections see \cite{Truskinovsky2}.
Macro and micro-level models for the reversible decohesion of finite elastic layers, associated with the denaturation of DNA, are considered in \cite{Truskinovsky1}.

In contrast to the study of dynamic fracture in mass-spring systems, very few articles concentrate on the failure of systems incorporating flexural elements such as beams.
The study of bending modes in 2D discrete flexural structures, composed of beams connecting masses and containing static faults, can be found in \cite{LSRyvkin}. The analysis of a static Mode III crack in a three-dimensional beam lattice, representing an open cell foam, is presented in \cite{Ryvkin1}. The investigation  of the fracture toughness of structured  materials, with applications to understanding the response of foams with cracks and broken elements, is considered in \cite{Ryvkin2}. The dynamics of periodic materials with small suddenly appearing flaws have been analysed in \cite{Ryvkin3}. 
Quasi-static damage propagation in two-dimensional beam structures under tensile loading has been considered in \cite{Cherkaev1}, motivating the design of fault-tolerant beam lattices in \cite{Cherkaev2}.

One-dimensional mass-beam chains supported by an elastic foundation, have been used to create a simplified model of a collapsing bridge \cite{BMS}.
This model has been further developed in \cite{BGMS} and used to predict the collapse rate of the San-Saba rail road bridge, Texas, in 2013. Transition waves in continuous flexural systems  have been shown to exhibit surprising non-steady failure  behaviour \cite{SASM}.

Experimentally, one can observe counter-intuitive fracture behaviour in elastic materials such as  rubber sheets \cite{MDS}, where analytical descriptions and numerical simulations of this process have been given in \cite{M2}. The influence of lattice vibrations and wave radiation processes on the stability of crack trajectories in Silicon crystals has been investigated experimentally and numerically using molecular dynamics simulations in \cite{Sherman1, Sherman2, Sherman3}. 

In \cite{NMS}, an analytical model for the failure of a discrete flexural structure, contained within an interface, was studied. The fracture was assumed to be caused by a remote sinusoidal load and the dynamic features of the structure
were studied. 
Further, in \cite{NMS2} the transient failure process was numerically modelled and the results of  \cite{NMS} were verified.

\begin{figure}[tp]
\centering
\includegraphics[width=14cm]{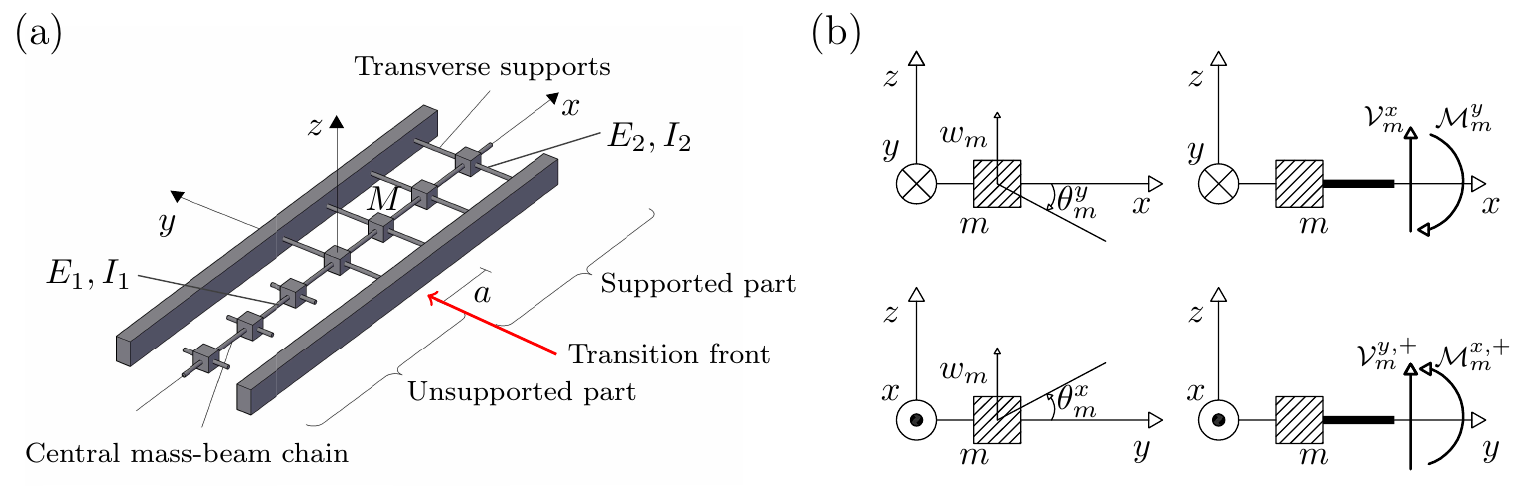}
\caption[]{{(a) A heterogeneous discrete structure composed of point masses connected by beams.  (b)  The convention adopted  for positive direction of the displacement $w_m$ and  rotations $\theta^{x}_m$ and $\theta^{y}_m$ associated with the $m^{th}$ mass. 
The positive directions for the internal bending moments ${\cal {M}}^{y}_m$, ($\mathcal{M}_m^{x, +}$)  and shear forces ${\cal {V}}^{x}_m$, ($\mathcal{V}^{y, +}_m$) in the $m^{th}$ beam directed from the $m^{th}$ mass in the positive $x$-direction ($y$-direction) are also indicated.} 
}
\label{Figura1}
\end{figure}

{Here, we consider  failure propagation   within the mass-beam structure shown in Figure \ref{Figura1}. The failure is represented by the sequential removal of transverse links supporting the masses along the  central chain, aligned with the horizontal axis. In \cite{NMS, NMS2}, the failure process was investigated under the assumption that the supporting columns of the central chain are removed if the masses  achieve a positive critical displacement. The steady-state regimes achieved in this case are referred to here as ``\emph{pure steady-state regimes}'' and assumed to be driven by a remote mechanical sinusoidal load.  In terms of applications, it is realistic to consider the case when the supporting links break  if a positive or negative critical displacement is reached. As we show here, along with the \emph{pure steady-state regimes} observed in \cite{NMS, NMS2}, the imposed fracture criterion yields additional regimes, which can be predicted and verified numerically. These regimes correspond to the case when, at each consecutive failure in the system, 
a change in the sign of the bending moments and shear forces in the supporting beams  can be observed. These regimes are investigated in detail here and are referred to as ``\emph{alternating generalised strain regimes}''. We note that for mass-spring systems, the analogous phenomenon is studied in \cite{MMS1}.}

The structure of the present article is as follows. In Section \ref{sec2}, we present an analytical description of the problem concerning failure propagation in the mass-beam structure subjected to a sinusoidal load. This section also includes the reduction of the problem to a functional equation of the Wiener-Hopf type, 
from which the alternating generalised strain regimes can be identified. The characterisation of the dispersive nature of this particular discrete system is given in Section \ref{secDISP}. The results of Section \ref{secDISP} are then used in Section \ref{sec4} to solve the Wiener-Hopf equation. We also identify the dynamic properties of the  system during the steady propagation of the alternating generalised strain regime in Section \ref{sec5}. In Section \ref{numsim}, we present numerical simulations which support the analytical results of Sections \ref{sec2}--\ref{sec5}. We give some conclusions in  Section \ref{Conclusions}. Finally, Appendices A and B contain details of some derivations.

\section{Model of failure within a discrete periodic flexural structure} 
\label{sec2}
\subsection{Description of the problem}\label{subsec2}

We consider a structure composed of a mass-beam chain, as shown in Figure \ref{Figura1}(a). This structure is formed from periodically placed point masses, connected longitudinally (along $x$-axis) by massless Euler-Bernoulli beams all with Young's modulus $E_1$ and second moment of area $I_1$. 
Each junction node has mass $M$ and corresponds to an index $m\in \mathbb{Z}$. This chain is assumed to be partially supported by transverse Euler-Bernoulli beams (parallel to the  $y$-axis), having Young's modulus $E_2$ and second moment of area $I_2$. These beams connect the masses to an interface where the beams are clamped. All beams  have length $a$.

Inside the structure, failure is assumed to propagate with a uniform speed $V$ as result of the breakage of the transverse connections to the interfaces within the structure. This breakage occurs when the  \emph{absolute value} of the displacement $w_p$ of the $p^{\textmd{th}}$ mass (where $m=p$ corresponds to the position of the  transition front in the structure, see Figure \ref{Figura1}(a), that depends on time) reaches a critical value $w_c$. Thus we assume for bonds to remain intact 
\begin{equation}\label{eq1}
|w_j(t)|<w_c, \quad j\ge p,  \quad p\in \mathbb{Z}\;,
\end{equation}
 and when the condition $w_p=\pm w_c$ is fulfilled, the transverse links at the $p^{th}$-mass break and the moving interface advances a distance $a$ in the structure to the $(p+1)^{\textmd{th}}$ node.

The condition (\ref{eq1}) 
enables us to exclude the case of non-steady propagation of the interface inside the structure.
Under these assumptions, we consider  the case when the \emph{generalised strains} (moments and shear forces) inside the transverse links alternate in sign during the failure process.

\subsection{Governing equations}\label{sec2.2}
{At a given time $t$, the position of the interface  in the structure is given by $m =\lfloor Vt/a \rfloor$, where $ \lfloor x \rfloor$ denotes the integer part of $x$. 
Here, the inequality $m \ge \lfloor Vt/a \rfloor$ corresponds to masses located in the supported part of the structure and $m <\lfloor Vt/a\rfloor$ are those in the unsupported region of the structure (see Figure \ref{Figura1}).}

The equations for the balance of linear and angular momentum for the $m^{th}$ mass, $m\in \mathbb{Z}$, are\begin{equation}\label{eqa}
M \frac{\D^2 w_m(t)}{\D t^2}= \CV^x_{m}(0,t)-\CV^x_{m-1}(a,t)+H(x-Vt)(\CV^{y, +}_{m}(0,t)-\CV^{y, -}_{m}(a,t))+Q_m(t),
\end{equation}
\begin{equation}\label{eqb}
\CM^y_{m}(0,t)=\CM^y_{m}(a,t),
\end{equation}
where $H$ is the Heaviside function
\[H(x)=\left\{ \begin{array}{ll} 1\;, & \quad  \mbox{ if } x\ge 0\;,\\
0\;,  &\quad  \mbox{otherwise}\;.
\end{array}\right.\]
In (\ref{eqa}) and (\ref{eqb}),   $\CM^y_m(\tilde{x},t)$ and $\CV^x_m(\tilde{x}, t)$,   denote the $y$-component of the  moment vector and the shear forces  in the $m^{th}$ horizontal  beam, respectively, at time $t$ {and position $\tilde{x}$ along the beam},  where $0<\tilde{x}<a$, (see Figure \ref{Figura1}(b)). They can be written in terms of the generalised coordinates describing the motion of the $m^{th}$ mass as
\begin{equation}
\label{eq1My}
\CM_m^y(\tilde{x}, t)=-\frac{2E_1I_1}{a^3}[a^2\theta^y_m(t)+a(a-3\tilde{x})(\theta_m^y(t)+\theta_{m+1}^y(t))-3(2\tilde{x}-a)(w_{m+1}(t)-w_m(t))]\;,\end{equation}
\begin{equation}
\label{eq1Vy}
\CV^{x}_m(\tilde{x}, t)=\frac{6E_1I_1}{a^3}[a(\theta_m^y(t)+\theta_{m+1}^y(t))-2(w_{m}(t)-w_{m+1}(t))]\;,
\end{equation}
where $w_m$ is  the displacement  and $\theta_m^y$ is the rotation about the $y$-axis of the $m^{th}$ mass (see Appendix A for the derivation of (\ref{eq1My}) and (\ref{eq1Vy})).
In a similar way, the shear forces $\CV^{y, +}_{j}(y_+,t)$ and  $\CV^{y, -}_{j}(y_-,t)$ correspond to those in the transverse beam located at  $0<y<a$ and $-a<y<0$, respectively, for $x=j a$ where  $j\ge \lfloor Vt/a\rfloor$. {Here, the notation $\tilde{y}_+=y$ ($\tilde{y}_-=y+a$) represents the local coordinate for the transverse beam located at $y>0$ ($y<0$)}. The shear forces $\CV^{y, -}_m$ adopt the same convention as described for $\CV^{y, +}_m$ in Fig. 2. As discussed in Appendix A, {the internal shear forces in these transverse elements} take the form
\begin{equation}\label{TF}
\CV_m^{y, \pm}(y_{\pm}, t)=\mp \frac{12E_2I_2}{a^3} w_m(t)\;.
\end{equation}
In (\ref{eqa}),  the term $Q_m(t)$ is the applied load at a node $m$. 
In addition, in Appendix A, we show the angular momentum balance about the $x$-axis for a mass in the supported region shows that the masses do not rotate about about the $x$-axis, i.e. $\theta^x_m=0$ for $m \in \mathbb{Z}$.

Equations (\ref{eqa}) and (\ref{eqb}) then become
\begin{multline}\label{eq1geAAA}
{6}\left\{2[2{w}_m(t) -{w}_{m-1}(t)-{w}_{m+1}(t)]-a[{\theta}^y_{m+1}(t)-{\theta}^y_{m-1}(t)]\right\} \\
+24r {w}_m(t)H(m-Vt/a)+\frac{Ma^3}{E_1 I_1}\f{\D^2{w}_m(t)}{\D t^2}=\frac{a^3}{E_1 I_1}Q_m(t)
\end{multline}
and
\begin{equation}\label{eq2geBBB}
3[{w}_{m+1}(t)-{w}_{m-1}(t)]+a[{\theta}^y_{m+1}(t)+{\theta}^y_{m-1}(t)+4{\theta}^y_m(t)]=0\,,
\end{equation}
where the contrast parameter $r=E_2I_2/E_1 I_1$. In (\ref{eq1geAAA}) and (\ref{eq2geBBB}), we  introduce the normalisations 
\[\tilde{t}={t}\sqrt{\frac{E_1I_1}{Ma^3}}\;,\quad  \tilde{v}=V\sqrt{\frac{Ma}{E_1 I_1}}, \quad \tilde{w}_m(\tilde{t})=\frac{{w}_m(t)}{a},\quad \tilde{Q}_m(\tilde{t})=\frac{a^2}{E_1 I_1}{Q_m}(t),\]
where the symbol ``tilde"  will be omitted in the following for ease of notation. In this case, the dimensionless governing equations for the system are then
\begin{multline}\label{eq1ge}
{6}\left\{2[2{w}_m(t) -{w}_{m-1}(t)-{w}_{m+1}(t)]-[{\theta}^y_{m+1}(t)-{\theta}^y_{m-1}(t)]\right\} \\
+24r {w}_m(t)H(m-vt)+\f{\D^2{w}_m(t)}{\D t^2}=Q_m(t)\,,
\end{multline}
and
\begin{equation}\label{eq2ge}
3[{w}_{m+1}(t)-{w}_{m-1}(t)]+[{\theta}^y_{m+1}(t)+{\theta}^y_{m-1}(t)+4{\theta}^y_m(t)]=0\,.
\end{equation}

\subsection{Derivation of the Wiener-Hopf equation}\label{Derivation}
Since we consider the case of alternating generalised strains,
we look for  the displacements and the rotations as functions of the moving coordinate~$\eta =m-vt$ as follows
\begin{equation}\label{assumption}
w_m(t)=(-1)^mw(\eta), \quad \theta^y(t)=(-1)^m\theta^y(\eta)\;,
\end{equation} 
where $w(\eta)$ and $\theta(\eta)$ play the role of envelope functions (see \cite{MMS1}). In addition, we assume the load takes the form   $Q_m(t)=(-1)^mQ(\eta)$.

We introduce the Fourier transforms with respect to the variable $\eta$ for the quantities $w^{{\rm F}}$, $\theta^{y{\rm F}}$  as
\begin{equation*}
\{w^{{\rm F}}, \theta^{y{\rm F}}\}=\int^\infty_{-\infty} \{w(\eta), \theta^y(\eta)\} e^{\text{i}{k}\eta}d\eta\;,
\end{equation*}
where $k$ is the dimensionless wavenumber.
The following ``half" transforms are also used:
\[ {{w}_{\pm}({k})}=\int^\infty_{-\infty} {w}(\eta) e^{\text{i}{k}\eta}H(\pm \eta)d\eta\;, \quad \pm \text{Im }k>0\;, \qquad w^{{\rm F}}=w_++w_-\;.\]
Here $w_+$ ($w_-$) corresponds to  a function analytic in the upper (lower) half of the complex plane defined by $k$.

After substitution of (\ref{assumption}) into (\ref{eq1ge}) and (\ref{eq2ge}) we obtain
\begin{multline}\label{eq3ge}
{6}\left\{2[2{w}(\eta) +{w}(\eta-1)+{w}(\eta+1)]+[{\theta}^y(\eta+1)-{\theta}^y(\eta-1)]\right\} \\
+24r {w}(\eta)H(\eta)+v^2\f{\D^2{w}(\eta)}{\D \eta^2}=Q(\eta)
\end{multline}
and
\begin{equation}\label{eq4ge}
-3[{w}(\eta+1)-{w}(\eta-1)]+[4{\theta}^y(\eta)-{\theta}^y(\eta+1)-{\theta}^y(\eta-1)]=0.\,
\end{equation}

The Fourier transform of (\ref{eq4ge})  with respect to the moving coordinate $\eta$ leads to 
\begin{equation}
{\theta}^{y{\rm F}}=-\frac{3\I \sin(k)}{2-\cos(k)}w^{{\rm F}}
\end{equation}
and together with the Fourier transform of (\ref{eq3ge}) we then obtain
\begin{eqnarray}\label{WH2}
&&h_1(k, 0+\text{i} k v)w_++h_2(k,  0+\text{i} k v)w_-{=Q^{\rm F}\;,}
\end{eqnarray}
where $Q^{\rm F}$ is the Fourier transform of the load $Q$ and
\begin{eqnarray}\label{eqdp1a}
&&h_j(k, Y)=\Omega^2_j(k+\pi)+Y^2\;, \quad j=1,2,\\
&&\Omega_1(k)=\sqrt{\frac{48\sin^4(k/2)}{2+\cos(k)}+24 r}\;, \quad \quad \Omega_2(k)=\sqrt{\frac{48\sin^4(k/2)}{2+\cos(k)}}\;.\label{neweq1a}
\end{eqnarray}
Here 
\begin{equation}\label{eqprelim}
 0+{\text{i}}kv=\lim_{\varepsilon\to +0} (\varepsilon+{\rm i}kv)
\end{equation}
 in (\ref{WH2}) appears as a result of the causality principle discussed in \cite{LSbook}, and represents the passing from the transient regime to the steady-state regime in the Laplace transform.
Note that the expressions  (\ref{neweq1a}) correspond to dispersion relations for the problem in \cite{NMS, NMS2}, where the fracture was considered for the pure steady-state failure regimes.

\section[]{Dispersion relations {and characterisation of waves in the structure}}\label{secDISP}
The dispersive properties of the structure can be determined from the zeros of $h_j$, $j=1,2,$ in  (\ref{WH2}). In replacing $Y$ by ${\rm i}\omega$ in (\ref{eqdp1a}) and rearranging for $\omega$ we obtain the dispersion relations 
\begin{eqnarray}\label{eqdp2}
\omega_1(k)=\Omega_1(k+\pi)=\sqrt{\frac{48\cos^4(k/2)}{2-\cos(k)}+24 r}\;, 
\end{eqnarray}
for the waves ahead of the transition front and
\begin{eqnarray}
\label{eqdp3}
\omega_2(k)=\Omega_2(k+\pi)=\sqrt{\frac{48\cos^4(k/2)}{2-\cos(k)}}\;,
\end{eqnarray}
for waves behind the transition front.
\begin{figure}[tp]
\centering
\includegraphics[width=10cm]{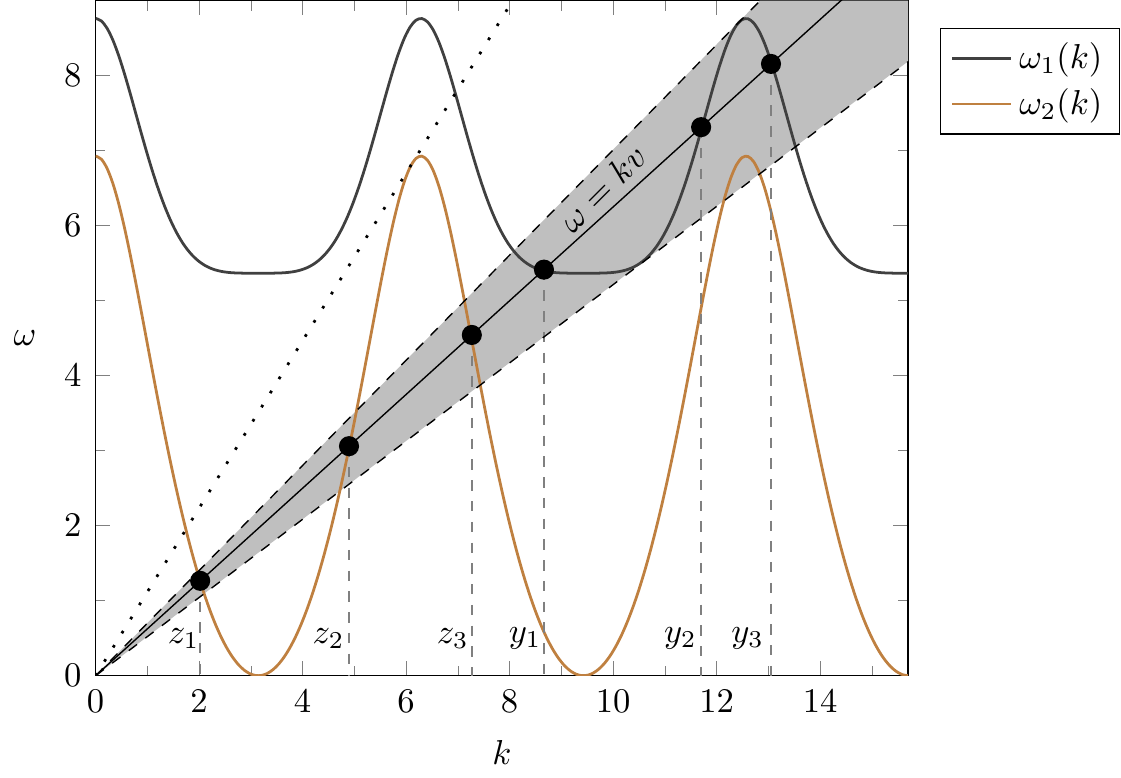}
\caption[]{
Dispersion diagrams for equations (\ref{eqdp2}) and (\ref{eqdp3}) are shown as functions of the wavenumber $k$ { for  $r=1.2$}. The ray $\omega=kv$ intersects the dispersion curves at points, shown as dots, with the  $k$-coordinates $y_1$, $y_2$, $y_3$  and $z_1$, $z_2$, $z_3$ that are solutions of $h_1$ and $h_2$, respectively. {The dotted inclined line has the gradient $v_{1}=1.117$ and  represents the upper bound of the alternating generalised strain failure regime. The dashed inclined lines represent the bounds for the speeds of possible transmission regimes, where waves propagating ahead of the transition front can occur. The occurrence  of such regimes can increase with decrease of the failure speed. Outside these intervals, failure regimes where evanescent waves propagate ahead of the transition front are encountered.}
}
\label{Figure00}
\end{figure}

Waves inside the structure are assumed to propagate as a result of either the external action from a remote load or through dissipation from the transition front.
These waves can be identified from the functions in \eqref{eqdp1a}. 
In Figure \ref{Figure00}, we give a representative example  showing the dispersion curves $\omega_1$ and $\omega_2$ as functions of the wave number $k$, based on \eqref{eqdp2}--\eqref{eqdp3}, along with the {generic} line
$\omega=kv$.

{{\bf \emph{Characterisation of the waves.}}\rm} Intersections of the line defined by $\omega=kv$  with the curve based on the relation $\omega_1$  in (\ref{eqdp2}) represent waves propagating ahead of the transition front ($\eta \ge 0$). Those intersection points of this line with $\omega_2$, given by (\ref{eqdp3}),  correspond to waves behind the transition front ($\eta<0$).

By computing the group velocity of the wave defined by $v_g=d\omega /dk$ at each intersection point, it is possible to determine the direction of propagation of the wave. Any intersection point of  the curve given by $\omega_1$  with the line $\omega=kv$  where $v_g<v$ or $v_g>v$, indicates a wave propagating in the supported region towards  or away from the  transition front, respectively.
In a similar way,  any intersection of the curve provided by $\omega_2$  with the ray $\omega=kv$ where $v_g>v$ or $v_g<v$ indicates a wave propagating in the unsupported region towards or away from the transition point, respectively.

{{\bf \emph{Roots of the functions $h_j$, $j=1, 2$.}}\rm}  Here we describe, in general, possible roots of the functions $h_j$, $ j =1, 2$, which will determine the nature of the waves inside the structure. These roots are associated with the intersections of the ray $\omega=kv$ and the curves determined by the functions $\omega_j$, $j=1,2$.

For the analytical model presented here, we consider the speed range 
$0<v \le {v_{1}=1.117}$. Here the upper bound determines the maximum speed for which  the failure process  can occur under the {{alternating generalised strain regime}}. This upper bound is the slope  of the dotted inclined line in Figure \ref{Figure00}.
For the failure process with speed $v$ to occur, it is required that there exists one intersection point of the line $\omega=kv$ with the curve $\omega_2$ where $v_g>v$. Such a point defines a wave that can propagate  behind  and deliver energy to the transition front.

{{\emph{Zeros of $h_1$.}}\rm} The function  $h_1(k, {\rm i}k v)$ has one, three or more pairs of simple zeros at the points $k=\pm y_{1}, \pm y_{2},   
\dots, \pm y_{2b+1}$, with $b\in \mathbb{Z}$, $b\ge 0$.
Here $b$ depends on the speed $v$. Prior to taking the limit in (\ref{eqprelim}), according to  \cite{LSbook}, these zeros are located in the complex plane and possess a small imaginary part. The sign of the imaginary part is then determined by comparing the group velocity $v_g$ with the failure speed $v$. For the points
\begin{enumerate}
\item with $k=\pm y_{1}, \dots, \pm y_{2b+1}$, the  group velocity $v_g<v$ and
\item with $k=\pm y_{2}, \dots, \pm y_{2b}$, the group velocity $v_g>v$.
\end{enumerate}
If $v_g<v$ ($v_g>v$), these points are located in the upper (lower) part of the complex plane defined by $k$. Thus in the  limit (\ref{eqprelim}), we have $k=\pm y_{1}+{\rm i}0, \dots, \pm y_{2b+1}+{\rm i}0$ and $k=\pm y_{2}-{\rm i}0, \dots, \pm y_{2b}-{\rm i}0$.

{Here, the points with $v_g<v$ can represent waves produced by a load situated far ahead of the transition front. In the loading problem considered in the next section, the points with $v_g>v$  will be associated with transmitted waves that propagate ahead of the transition front in the steady-state failure process. These waves can appear for particular failure speeds and an example of an interval of such speeds is shown {in the shaded region} in Figure \ref{Figure00}. In this case, {the  boundaries of this region correspond to the failure speed $v={0.5214}$} and $v={0.703}$.}

{{\emph{Zeros of $h_2$.}}\rm}
The function $h_2(k, {\rm i}k v )$ can have three, five or more pairs of  simple zeros at $k=\pm z_{1}, \pm z_{2}, \pm z_{3}, \dots,  \pm z_{2l+1}$, $l\in \mathbb{Z}$, $l\ge 1$. Again $l$ is a parameter which depends on the speed $v$. For
\begin{enumerate}
\item $k=\pm z_{1}, \pm z_{3}, \dots, \pm z_{2l+1}$, we have $v_g<v$,
\item $k= \pm z_{2},  \pm z_{4}, \dots, \pm z_{2l}$, it holds that $v_g>v$.
\end{enumerate}
Again, in  the limit in (\ref{eqprelim}), we receive $k=z_1+{\rm i}0, \dots,  z_{2l+1}+{\rm i}0$ and $k= \pm z_{2}-{\rm i}0,   \dots, \pm z_{2l}-{\rm i}0$.

{In the next section, in the  problem concerning the loading of the structure,  the wave numbers here with a negative or positive imaginary part will correspond to feeding or  reflected waves, respectively, that appear behind the transition front during the steady failure regime.}

\section{The Wiener-Hopf equation and fracture criterion
}\label{sec4}

Here, we  develop  the solution of (\ref{WH2}) for the alternating generalised strain regime.

We rewrite this  equation in the form
\begin{eqnarray}\label{WH1}
&& w_++\frac{\Psi_+(k)}{\Psi_-(k)}L(k)w_-=\frac{Q^{\rm F}}{h_1(k, 0+{\rm i}k v)}\;,
\end{eqnarray}
with 
\begin{eqnarray}
&&\Psi_+(k)=\frac{\displaystyle{\prod_{i=1}^{{l}}(0-\I (k-z_{2i}))(0-\I (k+z_{2i}))}}{\displaystyle{(1-\I k)^{2(l-b)}\prod_{j=1}^{{b}}(0-\I (k-y_{2j}))(0-\I (k+y_{2j}))}}\;,\label{def1}\\
&& \Psi_-(k)=\frac{\displaystyle{(1+\I k)^{2(l-b)}\prod_{j=0}^{{b}}(0+\I (k-y_{2j+1}))(0+\I (k+y_{2j+1}))}}{\displaystyle{\prod_{{i=0}}^{{l}}(0+\I (k-z_{2i+1}))(0+\I (k+z_{2i+1}))}}\;,
\end{eqnarray}
where $\prod_{j=1}^0=1$ and
\begin{eqnarray}\label{defL}
 \quad  L(k)= \frac{\Psi_-(k)}{\Psi_+(k)}\frac{h_2(k, 0+{\rm i}k v)}{h_1(k, 0+{\rm i}k v)}\;.
\end{eqnarray}
Here, $L(k)>0$ for $k\in \mathbb{R}$ and satisfies the conditions required for its factorisation. That is,
for $k\in \mathbb{R}$
\begin{eqnarray}
\text{Re}(L(k))=\text{Re}(L(-k))\quad \text{ and } \quad \text{Im}(L(k))=\text{Im}(L(-k))\;.
\end{eqnarray}

In addition,
$L(k)\to 1$ as $k\to \pm \infty$ and the index of the function $L(k)$ is zero. Then, $L(k)$ can be written using the Cauchy-type factorisation in the form:
\[L(k)=L_+(k)L_-(k)\;, \quad L_\pm(k)=\text{exp}\left(\pm \frac{1}{2\pi \I}\int_{-\infty}^\infty \frac{\ln L(\xi)}{\xi-k}\, d\xi \right)\;, \quad  \pm \text{Im}(k)>0\;.\]
In this representation,  the function $L_+$  ($L_-$) is analytic in the upper (lower) half of the complex plane defined by $k$. 
Then, (\ref{WH1}) can be written in the form of the Wiener-Hopf equation
\begin{equation}\label{WHEQM}
\frac{1}{L_+(k){\Psi_+(k)}}w_++\frac{1}{\Psi_-(k)}L_-(k)w_-=\frac{Q^{\rm F}}{L_+(k)\Psi_+(k)h_1(k, 0+{\rm i}k v)}.
\end{equation}
\subsection{The solution of the Wiener-Hopf equation}
As in \cite{LSbook}, non-trivial solutions of (\ref{WHEQM}) correspond to singular points of the right-hand side. Such singular points occur when 
\[{L_+(k)\Psi_+(k)h_1(k, 0+{\rm i}k v)}=0\;.\]
 As $L(k)$ has no real zeros and, referring to Section \ref{secDISP} and  (\ref{def1}), we note the above left-hand side is only zero when
$k=\pm y_{2j+1}-\I 0$, $0\le j \le b$, and $k=\pm z_{2j}-\I 0$, $1\le j \le l$.  The points $k=\pm y_{2j+1}-\I 0$, $0\le j \le b$, correspond to the action of a remote force  ahead of the transition front as they are associated with the inequality $v_g<v$. On the other hand, the points  $k=\pm z_{2j}-\I 0$, $1\le j \le l$, represent the loading from behind the transition front (with $v_g>v$).

Here, we assume that the loading of the structure takes the form of an oscillatory mechanical load. This load is situated {at some point far from the transition front on the negative $x$-axis}
with frequency $\omega_0=vk_f$, where we note  in addition $\omega_0=\omega_1(k_f)$. Here   $k_f$, corresponds to a feeding wave that causes the failure to propagate with constant speed $v$ (see section \ref{FW1.1} for the description of how $k_f$ is chosen).  Following \cite{LSbook}, this allows one to rewrite the right-hand side of (\ref{WHEQM}) as
\begin{eqnarray}
\frac{{1}}{L_+(k)\Psi_+(k)}w_++\frac{L_-(k)}{\Psi_-(k)}w_-&=&\Big[ \frac{Ce^{\I \phi}}{0-\I (k-k_f)}+  \frac{\overline{C}e^{-\I \phi}}{0-\I (k+k_f)}\nonumber \\
&&+ \frac{{C}e^{\I \phi}}{0+\I (k-k_f)}+  \frac{\overline{C}e^{-\I \phi}}{0+\I (k+k_f)}\Big], 
\end{eqnarray}
where $\phi$  is the phase of the feeding wave and $C$ is a constant to be linked to the amplitude of the feeding wave.
The solution to the above equation is then found in the form
\begin{eqnarray}
&&w_+=L_+(k)\Psi_+(k)\left[ \frac{Ce^{\I \phi}}{0-\I (k-k_f)}+  \frac{\overline{C}e^{-\I \phi}}{0-\I (k+k_f)}\right]\;,\label{solp}\\
&& w_-=\frac{\Psi_-(k)}{L_-(k)} \left[ \frac{Ce^{\I \phi}}{0+\I (k-k_f)}+  \frac{\overline{C}e^{-\I \phi}}{0+\I (k+k_f)}\right]\;.\label{solm} 
\end{eqnarray}

\subsection{The fracture criterion and uniqueness of the solution}
In terms of the moving coordinate system, the condition (\ref{eq1})
can be re-interpreted for the amplitude function $w(\eta)$ as
\begin{equation}\label{wceta}
w(0)=w_c, \quad  \text{ and } \quad w^\prime (+0)<0\;.
\end{equation}
The first of these conditions can be satisfied by computing the limits 
 \begin{equation}\label{eqr}
 w(\eta)\big|_{\eta=0}=\lim_{k \to \I \infty}-\I k\, w_+=\lim_{k \to -\I \infty}\I k\, w_-
 \end{equation}
 using (\ref{solp}) and (\ref{solm}).
In doing this 
we obtain
\begin{equation}\label{wc1}
w_c=2 \text{Re}(C e^{\I \phi})\;,
\end{equation} 
which allows for the determination of $\phi$ if the feeding wave amplitude, and consequently $C$, are known.
With regard to the second condition in (\ref{wceta}), we construct the asymptote of $w_+$ in (\ref{solp}) as $k\to \I \infty$. which in terms of the physical variables represents the expansion
\[w(\eta)=w(+0)+\eta w^\prime(+0)+O(\eta^2)\quad \text{ for } \eta \to +0\;.\]
We have 
\[w_+=-\frac{2\text{Re}(C e^{{\rm i}\phi})}{{\rm i}k}-\frac{2}{k^2}\left[k_f \text{Im}(C e^{{\rm i}\phi} )+\text{Re}(C e^{{\rm i} \phi})\left\{2(b-l)+\frac{1}{2\pi}\int^\infty_{-\infty} \ln L(\xi)d\xi\right\}\right]+O\left(\frac{1}{k^3}\right)\]
for $k\to {\rm i} \infty$. Owing to the fact that 
\[\int_0^\infty \eta e^{{\rm i}k \eta} dk =\frac{1}{(0-{\rm i} k)^2}\;,\]
we obtain  that the second condition in  (\ref{wceta}) is satisfied if
\begin{equation}\label{phidet}
k_f\text{Im}(C e^{{\rm i} \phi})+\text{Re}(C e^{{\rm i} \phi})\left\{2(b-l)+\frac{1}{2\pi}\int^\infty_{-\infty}\ln  L(\xi)d\xi\right\}<0\;.
\end{equation}
This allows for  $\phi$ to be uniquely determined as a solution of (\ref{wc1}).

\section{Dynamic loading and waves in the steady-state fracture regimes}\label{sec5}
We discuss the dynamic behaviour of the structure, subjected to  a sinusoidal load,  during the alternating generalised strain failure process. In particular,  concerning the wave radiation processes associated with the moving transition front, we discuss the reflected and transmitted waves emitted from this point.

\subsection{Sinusoidal remote load and  the feeding wave}\label{sec5.1}
We assume a sinusoidal load acts at some point far from the origin on the negative $x$-axis (see Figure \ref{Figura1})
inside the structure. The load has amplitude $P_0$ and frequency $\omega_0$.
Following the normalisations adopted in Section \ref{Derivation}, we introduce the normalised forms of these quantities as 
\[\tilde{\omega}_0=\sqrt{\frac{Ma^3}{E_1I_1}}{\omega_0}\;, \quad \tilde{P}_0=\sqrt{\frac{a^2}{E_1I_1}}{P}_0\;.\]
From here on, we omit  the symbol $``\,\tilde{\,\,}\,"$.
This load generates a 
wave incident on the moving interface with the form 
\begin{equation}\label{FWform}
w_f(\eta)=\frac{P_0}{2k_fR_0(k_f, \omega_0)}\cos(k_f \eta -\phi)\;,
\end{equation}  
where $\phi$ is the feeding wave phase, $\eta=m-\omega_0 t/k_f$, with $\omega_0/k_f=v$ being  the phase speed of the outgoing wave and
\begin{equation}
\label{R0def}
R_0(k, \omega_0)=\frac{h_2(k, \text{i}\omega_0 )}{(0-\I (k-k_f))(0+\I (k-k_f))}\;,
\end{equation}
 (see the Appendix B for the derivation of the amplitude in (\ref{FWform})).

We now determine the expression for the constant $C$ in the solution (\ref{solp}), (\ref{solm}) by considering the fracture criterion at the point $\eta=0$.
We note that the form of the feeding wave based on (\ref{FWform}) can also be found if one considers the complex residues of the simple poles at $k=\pm k_f$ for the  inverse Fourier transform of $w_-$. This is derived 
under the assumption $\eta \to -\infty$, corresponding to a remote distance far behind the transition front and using 
\begin{equation}
\frac{1}{2\pi}\int_{-\infty}^\infty w_-(k)e^{-\I k \eta}dk\;.
\end{equation} 
In doing so, we have
\begin{equation}
w_f=\Big[\frac{\Psi_-(k_f)}{L_-(k_f)}Ce^{\I (\phi-k_f\eta) }+\frac{\Psi_-(-k_f)}{L_-(-k_f)}\overline{C}e^{-\I (\phi-k_f\eta)}\Big]\;.
\end{equation}
Noting that 
\[\Psi_\pm (-k)=\overline{\Psi_{\pm}(k)}, \quad L_\pm (-k)=\overline{L_{\pm}(k)}\;,\]
we obtain
\begin{equation}
w_f=2\text{Re}\Big[\frac{\Psi_-(k_f)}{L_-(k_f)}Ce^{\I (\phi-k_f\eta) }\Big]\;.
\end{equation}
 Comparing with (\ref{FWform}), we see that
\begin{equation}\label{C}
C=\frac{L_-(k_f)P_0}{4k_fR_0(k_f, \omega_0)\Psi_-(k_f)}\;.
\end{equation}

\subsection{Criterion for steady-state failure propagation in the alternating generalised strain regime}\label{MinFW}

\begin{figure}[tp]
\centering 
\includegraphics[width=0.6\textwidth]{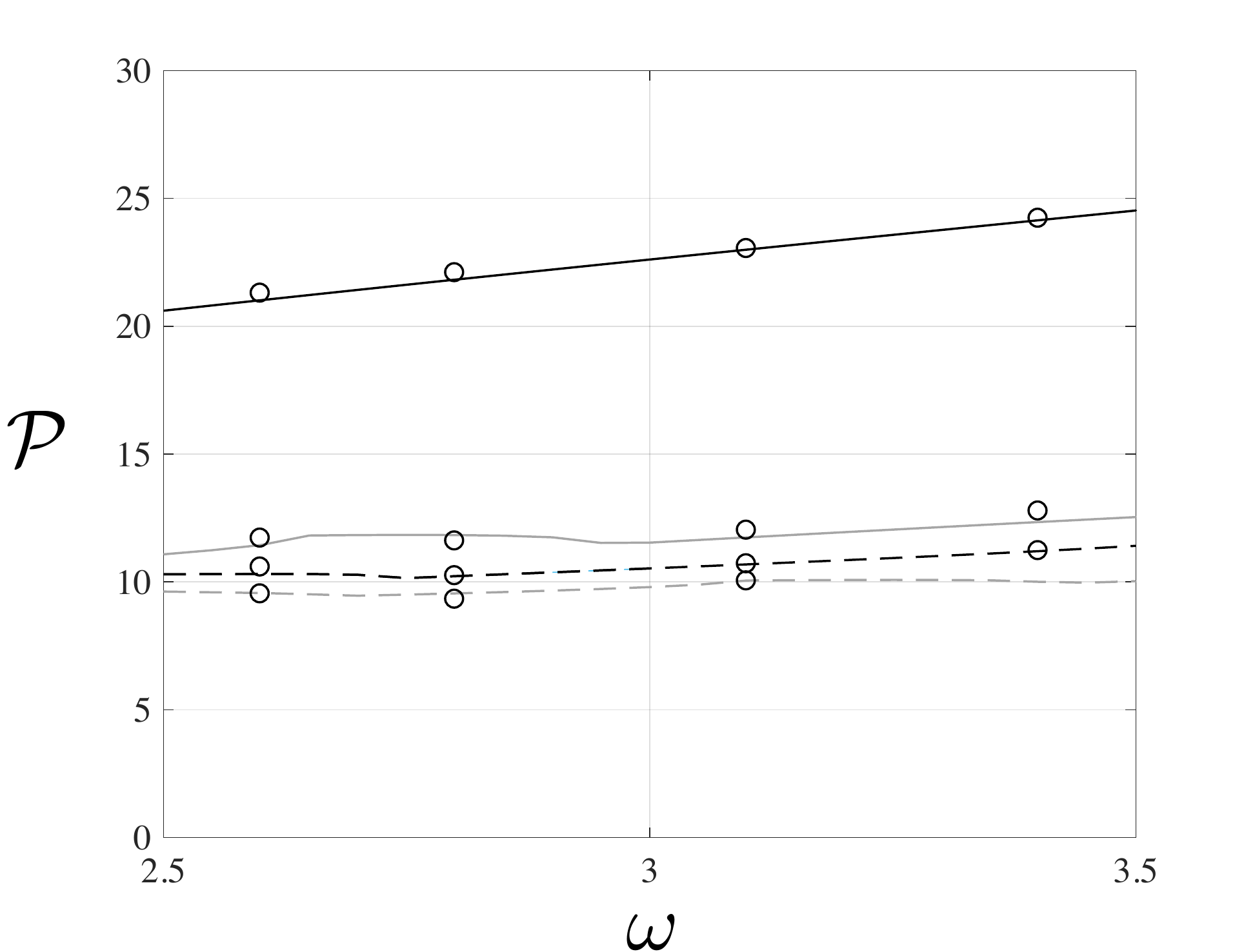}
\caption{{The plot of $\CP_{\text{min}}$ in a typical range of loading frequencies $\omega$. Black curves are associated with the pure steady-state regimes identified in \cite{NMS, NMS2} and the grey curves are computed using (\ref{minloadpred}) and (\ref{Pmin}). 
The circles are results obtained from numerical simulations implemented in MATLAB  (discussed in Section \ref{numsim}), where a sufficiently large system loaded by a sinusoidal load is considered and these failure regimes are identified.}
{For a given frequency in this interval, each curve provides the $\CP_{\text{min}}$ for the failure to propagate with a speed $v=\omega/k_f$. The speed is different for each curve and in this case the speeds form a monotonically increasing sequence, with the grey dashed curve associated with the lowest failure speed and the solid black curve corresponding to the highest possible failure speed.}}
\label{min_load}
\end{figure}

For the crack to propagate steadily, it is required that amplitude of the feeding wave reaches or exceeds the critical displacement, i.e.
\[w_f(-0)\ge w_c\;.\]
Using this condition, (\ref{wc1}) and (\ref{C}) yields the criterion
\begin{equation}\label{minloadpred}
\CP \ge \CP_{\text{min}}
\end{equation}
for the failure to propagate steadily in the alternating generalised strain regime, where
\begin{equation}\label{Pmin}
\CP=P_0/w_c \qquad \text{ and } \qquad  \CP_{\text{min}}:= \min \CP=\frac{2k_f R_0(k_f, \omega_0)|\Psi_-(k_f)|}{|L_-(k_f)|}\;.
\end{equation}
Figure \ref{min_load}, shows the plot of $\CP_{\text{min}}$ in (\ref{Pmin}) for a representative range of load frequencies. The computations are performed for $r=0.4$, corresponding to a structure with soft supports, which will be investigated further in Section \ref{numsim}.  For a given choice of the load frequency, the  grey solid and dashed lines, based on   (\ref{Pmin})  in Figure \ref{min_load}, predict the $\CP_{\text{min}}$ required for a fracture process with speed $v=\omega/k_f$ to propagate through the flexural system. Here $k_f$  is determined from the dispersion diagram. The grey  solid curve corresponds to regimes with a higher fracture speed than those connected with the grey dashed curve. These curves are associated with the $\CP_{\text{min}}$ for the alternating generalised strain regime.
 
For illustrative purposes, we  also plot the results of \cite{NMS, NMS2} with black curves and these predict the appearance of pure steady-state modes, where the sign of the bending moments and shear forces are uniform at each instant of the failure process.

Accompanying these analytical computations is a verification of the predictions based on numerical simulations for a sufficiently large finite structure implemented in MATLAB (and described in detail in Section \ref{numsim}). 
There is a excellent agreement between the analytical results, which are based on an infinitely long medium, and the results of the MATLAB computations {based on a finite medium}.

\subsection{Dynamic features for the alternating generalised strain regime}\label{AGSR}

Here, we give the representation of the envelope functions for the reflected and transmitted waves encountered during the failure process based on \eqref{solp} and \eqref{solm}. The expressions for the associated waveforms in the structure may then be determined by combining the results presented here with \eqref{assumption}.

{{\emph{Reflected waves behind the transition front.}}\rm}
The reflected waves  can be derived from the residues of the poles $k=\pm z_{2j+1}+\I0$, $1\le j \le  l$, of the function $w_-$ in \eqref{solm}. 
The functions $w_r^{(s)}$,  $0\le s \le  l$, representing the reflected waves, have the following form
\[w_r^{(s)}=A_r^{(s)} \cos(z_{2s+1}\eta -\psi_r^{(s)})\;,\]
with amplitude $A_r^{(s)}$ given by
\[A_r^{(s)}=\frac{4|C||\Psi_-^r(z_{2s+1})|}{ |L_-(z_{2s+1})||z_{2s+1}^2-k_f^2|}\sqrt{z_{2s+1}^2\cos^2(\phi+\psi_c)+k_f^2\sin^2(\phi+\psi_c)}\]
where
\[\Psi_-^r(z_{2s+1})=\frac{(-1)^{b-l+1} (1+\I z_{2s+1})^{2(l-b)}}{2 \I z_{2s+1}}\frac{\displaystyle{\prod_{j=0}^{b} (z_{2s+1}^2-y_{2j+1}^2)}}{\displaystyle{\prod_{\substack{0\le i \le l\\i\ne s}}(z_{2s+1}^2 -z_{2i+1}^2)}}\]
and the  phase shift $\psi_r^{(s)}$ given as
\[\psi_r^{(s)}=\text{arg}\Big(-\frac{\I\Psi_-^r(z_{2s+1})(z_{2s+1}\cos(\phi+\psi_c)+\I k_f \sin(\phi+\psi_c))}{(z_{2s+1}^2-k_f^2)L_-(z_{2s-1})}\Big)\;, \quad \psi_c=\text{arg}({C})\;.\]

{{\emph{Waves transmitted ahead of the transition front.}}\rm}
The waves that are transmitted ahead of the propagating interface are associated with the poles of the function $w_+$ in \eqref{solp}, at the points $k=y_{2j}-\I 0$, $1\le j \le b$.

The transmitted waves $w_{tr}^{(s)}$, $1\le s \le b$, are given by:
\[w_{tr}^{(s)}=A_{tr}^{(s)} \cos(y_{2s}\eta -\psi_{tr}^{(s)})\;.\]
Here the amplitude $A_{tr}^{(s)}$ has the form
\[A_{tr}^{(s)}=\frac{4|C||\Psi_+^{tr}(y_{2s})||L_+(y_{2s})|}{|y_{2s}^2-k_f^2|}\sqrt{y_{2s}^2\cos^2(\phi+\psi_c)+k_f^2\sin^2(\phi+\psi_c)}\]
with
\[\Psi_+^{tr}(y_{2s})=\frac{(-1)^{l-b+1}}{2 \I y_{2s} (1-\I y_{2s})^{2(l-b)}}\frac{\displaystyle{\prod_{j=1}^{l} (y_{2s}^2-z_{2j}^2)}}{\displaystyle{\prod_{\substack{1\le i \le b\\ i\ne s}}(y_{2s}^2 -y_{2i}^2)}}\;.\]
The phase shift $\psi_{tr}^{(s)}$ is 
\[\psi_{tr}^{(s)}=
\text{arg}\Big(\frac{\I {\Psi_+^{tr}(y_{2s})}{L_+(y_{2s})}[y_{2s}\cos(\phi+\psi_c)+\I k_f \sin(\phi+\psi_c)]}{y_{2s}^2-k_f^2}\Big)\;.\]
Note, if $b=0$, then there are no propagating waves transmitted to the intact structure.

\section{Numerical simulations modelling the failure process}\label{numsim}

In this section, we implement a numerical scheme  in MATLAB that models the behaviour of a finite flexural structure subjected to a sinusoidal load.  In addition, using the numerical scheme we also trace the propagation of failure  inside this structure  as a result of the action of the load. The scheme is based on the normalised governing equations (\ref{eq1ge}) and (\ref{eq2ge}) for the system,  which are solved using the \texttt{ode45} routine of MATLAB. 

We use the results from the analysis of the  alternating generalised strain fracture regime 
presented in Section \ref{sec2},  to show that we can predict: (i) when such regimes occur, (ii) the speed with which the fracture propagates in such regimes and (iii) the behaviour of the structure in these particular regimes.

\subsection{The numerical model}
A  finite structure composed of~3800 nodes,  connected by massless beams, is considered.  {As the numerical scheme is based on the dimensionless equations (\ref{eq1ge}) and (\ref{eq2ge}), the dimensionless length  and flexural stiffness of the connecting beams can be taken equal to unity}. Therefore, the position of each mass is then given by $x=n$, $1\le n \le  3800$. The system is initially at rest.    We consider this structure composed of a region supported by transverse links  and a region where the masses along the central axis of the structure are unsupported by such links. {The supported region is characterised by the heterogeneity parameter $r$}. 
The nodes corresponding to $0\le x\le  1999$ constitute the unsupported part of the structure, behind the interface, and the remaining~1800 nodes (corresponding to $2000\le x\le 3800$) represent the system in the supported region. Therefore, the interface between these two media is initially at $n=2000$.

{The external applied force is taken as a sinusoidal force having the form $P_0\sin({\omega}_0 t)$, where  $P_0$ and  ${\omega}_0$ are the dimensionless load amplitude and  frequency.}
This load is situated at node $x=1800$. We can choose the value of ${\omega}_0$ based on the frequency $\omega$ of the unforced  problem for the unsupported structure of infinite extent {in the $x$ direction}. {To generate feeding waves in the unsupported structure capable of reaching the transition front, the load frequency $\omega_0$ should be chosen in the interval $0<\omega_0<\sqrt{48}$, representing the passband for the analogous infinite medium. Below, we consider $\omega_0=3.1$ and 5.9 representing frequencies located  in the middle and close to the upper limit of the passband, producing waves capable of  exciting  the microstructure. In the numerical scheme, the right end of the structure is clamped at $x=3800$, whereas the node at $x=1$ is  free. The structure's size has been chosen to minimise effects due to  reflections produced by the ends.}

 As discussed in Section \ref{sec5.1}, the  feeding waves generated by the load help to initiate and propagate the failure process. Fracture regimes in two different types of structure are examined here:
 \newline
\\
 \underline{Case 1 
  }:
A structure whose supporting transverse links are~``softer'' than those  along the central axis of the structure ($r=0.4$)
\newline
\\
\underline{Case 2 
  }: A structure with transverse supports that are much stiffer than the beams along its central axis ($r=3.4$).

\subsection{Case 1 - soft supports}\label{FW1}
Here we show that it is possible to encounter two fracture regimes. The first involves the regime where the failure will propagate if \eqref{eq1} is violated at the transition front. The second concerns the regime defined by 
\begin{equation}\label{eq1a} 
w_j(t) <w_c, \quad j\ge p, \quad p, j\in \mathbb{Z}\;,
\end{equation}
where $p$ represents the node number where the interface is located at a given time $t$.
This was studied in detail in \cite{NMS, NMS2}.  For these regimes, the sign of the generalised strains at the instant of each fracture in the structure remains constant. {We refer to the regimes associated with (\ref{eq1a}) 
 as the \emph{pure steady-state regimes}}.

We first show that speeds associated with the failure regimes observed can be predicted from the dispersion curves~\eqref{eqdp2}, \eqref{eqdp3} and those identified in~\cite{NMS, NMS2}.
\begin{figure}[tp]
\centering
\includegraphics[width=14cm]{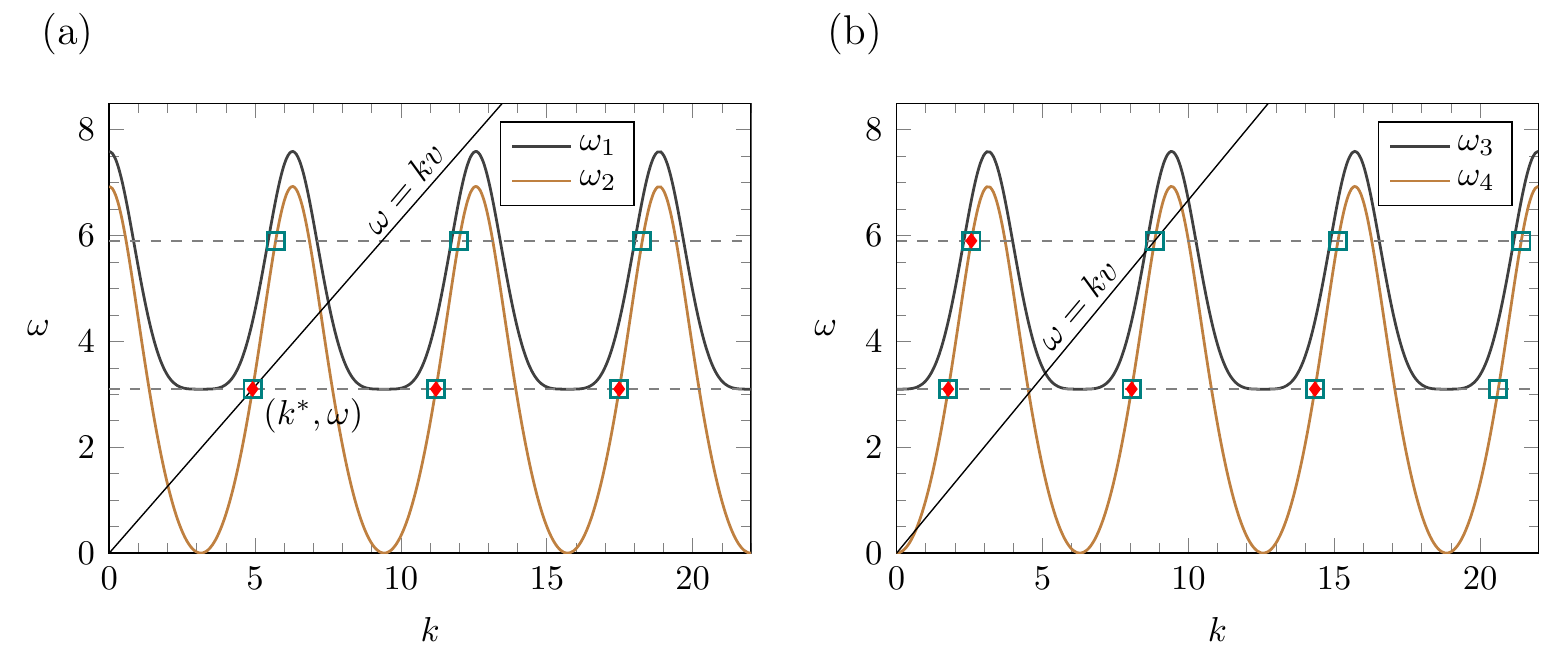}
\caption[]{Dispersion relations 
for~$r=0.4$. (a)  The dispersion curves $\omega_1$ and~$\omega_2$ for the case of the alternating generalised strain regime for fracture propagation {(see (\ref{eqdp2}) and (\ref{eqdp3})}). (b) The dispersion curves~$\omega_3$ and~$\omega_4$ for the  pure steady-state regimes that are associated with the conditions \eqref{eq1a} (see~\cite{NMS, NMS2}). The loading  frequencies~$\omega_0=3.1$ and~$5.9$ are shown as dashed horizontal lines. {The squares represent the intersection points of lower curves with these lines, where the wave group velocity $v_g$ exceeds the wave phase velocity $v$}. The slope of the ray~$\omega=kv$ that passes through the intersection points indicates possible steady-state speeds. Those intersection points with the red diamonds are associated with failure regimes observed in the MATLAB numerical simulations. (online version in colour)}
\label{DispersionCurves1}
\end{figure}

{{\emph{Failure speed predictions.}}\rm}\label{FW1.1}
A list of steady-state speeds for the considered fracture regimes can be determined using the dispersion curves discussed in Section \ref{secDISP}. Figure~\ref{DispersionCurves1}(a) shows the dispersion curves~\eqref{eqdp2}, \eqref{eqdp3} plotted as functions of the wave number~$k$ for the case~$r=0.4$. In addition, the horizontal dashed lines  represent the loading frequencies $\omega_0$ chosen for the MATLAB computations. 

For the case $\omega_0=3.1$,
the horizontal dashed line in  Figures \ref{DispersionCurves1}(a) and (b) associated with this frequency intersect the curves given by~$\omega_2(k)$ and~$\omega_4(k)$ infinitely many times. Since we require a feeding wave to reach the transition front inside the structure, we need the group velocity $v_g$
of this wave to be greater than the fracture speed~$v$ i.e. the slope of the line $\omega=kv$. For~$k>0$, we have indicated using squares the intersection points, representing possible feeding waves, where this criterion is satisfied in Figure~\ref{DispersionCurves1}(a) and (b). 

Without loss of generality, we take the first of these points $(k^*, \omega)$, shown in Figure \ref{DispersionCurves1}(a), along the line  $\omega=3.1$ with the smallest wavenumber. We connect this point to the origin
by a ray whose slope is defined by~$v=\omega/k^*$ (see Figure \ref{DispersionCurves1}(a)). This slope predicts a possible steady-state speed of failure inside the structure. In the case of~$\omega=3.1$ in Figure~\ref{DispersionCurves1}(a), such an intersection point is given by~$(k^*, \omega)=(4.915, 3.1)$ and the slope of the corresponding ray is~$0.6307$. That particular intersection point corresponds to an alternating generalised strain failure regime with speed $v=0.6307$ initiated by the sinusoidal load of frequency $\omega=3.1$.

Owing to the~$2\pi$-periodicity of the function~$\omega_2(k)$, a decreasing sequence of possible speeds can be deduced in the form
\[v=\frac{\omega}{k^*+2\pi n}\;, \quad n\ge 0, n\in \mathbb{Z}\;.\]
As an example, for the case~$\omega=3.1$ in Figure \ref{DispersionCurves1}(a), this list takes the form
\begin{equation}\label{list1}
v= {\text{0.6307, 0.2768, 0.1773, 0.1304, 0.1032, ...}}\;.
\end{equation}
The intersection points connected with these speeds are shown along the line defined by~$\omega=3.1$ in Figure~\ref{DispersionCurves1}(a) {by squares}. 

 In the MATLAB simulations we can also identify pure steady-state regimes corresponding  to the case when the  failure propagation occurs as a result of the violation of~\eqref{eq1a} at the transition front. In this scenario, following~\cite{NMS, NMS2}, the dispersion relations  are given by~
 $\omega_{j+2}(k)=\Omega_j(k)$, $j=1,2,$ (see \eqref{eqdp1a}) and are shown in Figure~\ref{DispersionCurves1}(b). Repeating the procedure outlined above, we can obtain the speeds in the 
  pure steady-state regime.
   For~$\omega=3.1$, the list of predicted speeds for this regime is:
\[v={\text{1.7484, 0.3848, 0.2162, 0.1503, 0.1152, ...}}\;.\]

Predictions for the potential steady-state speeds at different load frequencies can be calculated in a similar way.
\begin{figure}
\centering
\includegraphics[]{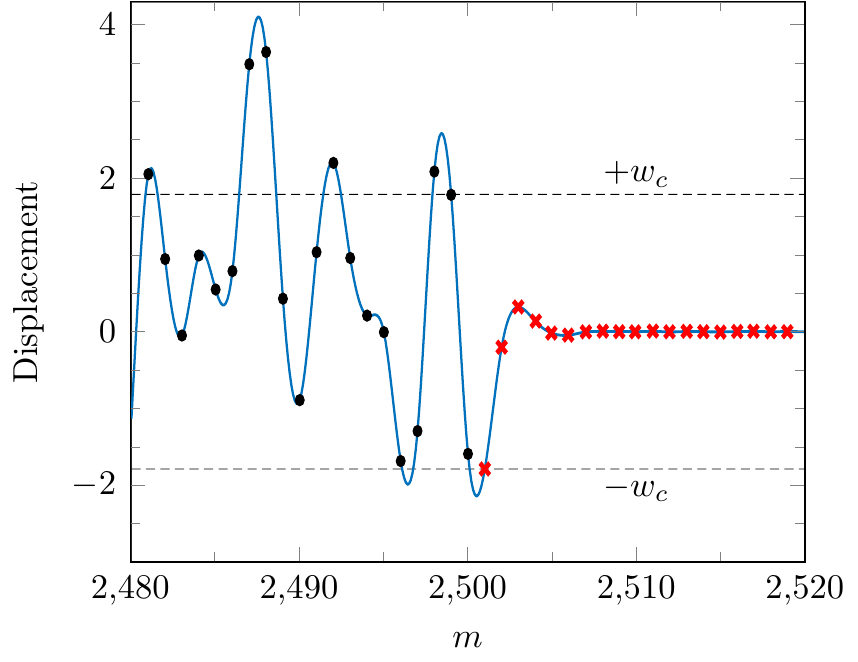}
\caption{Displacement profile attained by the  central axis of the  harmonically excited flexural system that is undergoing a steady failure with the alternating generalised strain mode. The frequency and amplitude of the external load is $\omega_0=3.1$ and $\CP=13.115$, respectively. The heterogeneity parameter of the structure $r=0.4$. The speed of failure for this particular regime is $0.6307$.   Masses along the unsupported structure are represented by black dots and those masses supported by transverse links  are marked with crosses. The limits $\pm w_c$, with $w_c=1.7919$, of the critical displacement for every mass in the supported region is indicated by horizontal dashed lines. (online version in colour)}
\label{fig_profile}
\end{figure}


{{\emph{Results of the transient analysis and identification of failure regimes.}}\rm}\label{subsec6}
Here, for $r=0.4$ and $\omega=3.1$, we analyse the failure processes in the finite structure for various load amplitudes. We show that the failure processes discussed here and in \cite{NMS, NMS2} both can appear as a result of the action of the load.
 
{Figure \ref{fig_profile} shows a result of the transient computation presented here when one of the alternating generalised strain failure regimes is initiated in the finite medium. As mentioned earlier, for a given load  frequency, the appearance of these regimes is dependent on the applied load amplitude. In Figure \ref{fig_profile}, we show the dynamic response of the structure when the speed of this failure regime is maximal. There, as predicted by the analytical results in Sections \ref{sec2}--\ref{sec5}, the transition front propagates with a constant speed as a result of feeding waves from the load providing energy to the front. This allows the front to reach the critical displacement for the failure of the supports. During the steady propagation, waves are radiated outward from the transition front when the failure of the supporting beams is achieved.  The video of the Supplementary data shows the evolution of this failure process and the behaviour of the structure resulting from the transient analysis undertaken here.
 Later we show that in the transient regime the failure process can occur non-steadily before reaching such states.  In particular, we will also demonstrate that dynamic response of the structured medium is different when undergoing an alternating generalised strain failure mode when compared with a pure steady failure mode.   }
 





\vspace{0.1in}\noindent {\bf \emph{Transient behaviour of the transition front} \rm}

\vspace{0.1in}\noindent {The failure of the finite structure resulting from the harmonic load was performed for a large interval of load amplitudes. We denote by $m^*$ the index of the mass where the transition front is initially located. For the simulations $m^*=2000$. The index $m_i=m-m^*+1$ refers to the index of a mass in the supported region  where the failure of the transverse beams at time $t_i$, $i \ge 1$, $i \in \mathbb{Z}$,  occurs. Simultaneously, as the length of the beams along the central axis is equal to unity, $m_i$ also represents the position of the failure in the supported region measured with respect to the initial position of the transition front.

 Figure \ref{Group1}(a) shows several examples of the failure position as a function of time $t$  for various load  amplitudes where steady failure regimes can be observed in the simulations. 
 Since the results are based on a transient simulation, we can expect some effects in the initial period of the failure process. For instance, for $\CP= 11.496$ in Figure \ref{Group1}(b) we see a visible oscillatory behaviour in the transition front speed initially.  Similar behaviour can also be seen for the computations based on the other load amplitudes. 
\begin{figure}[tp]
\centering
\includegraphics[width=14cm]{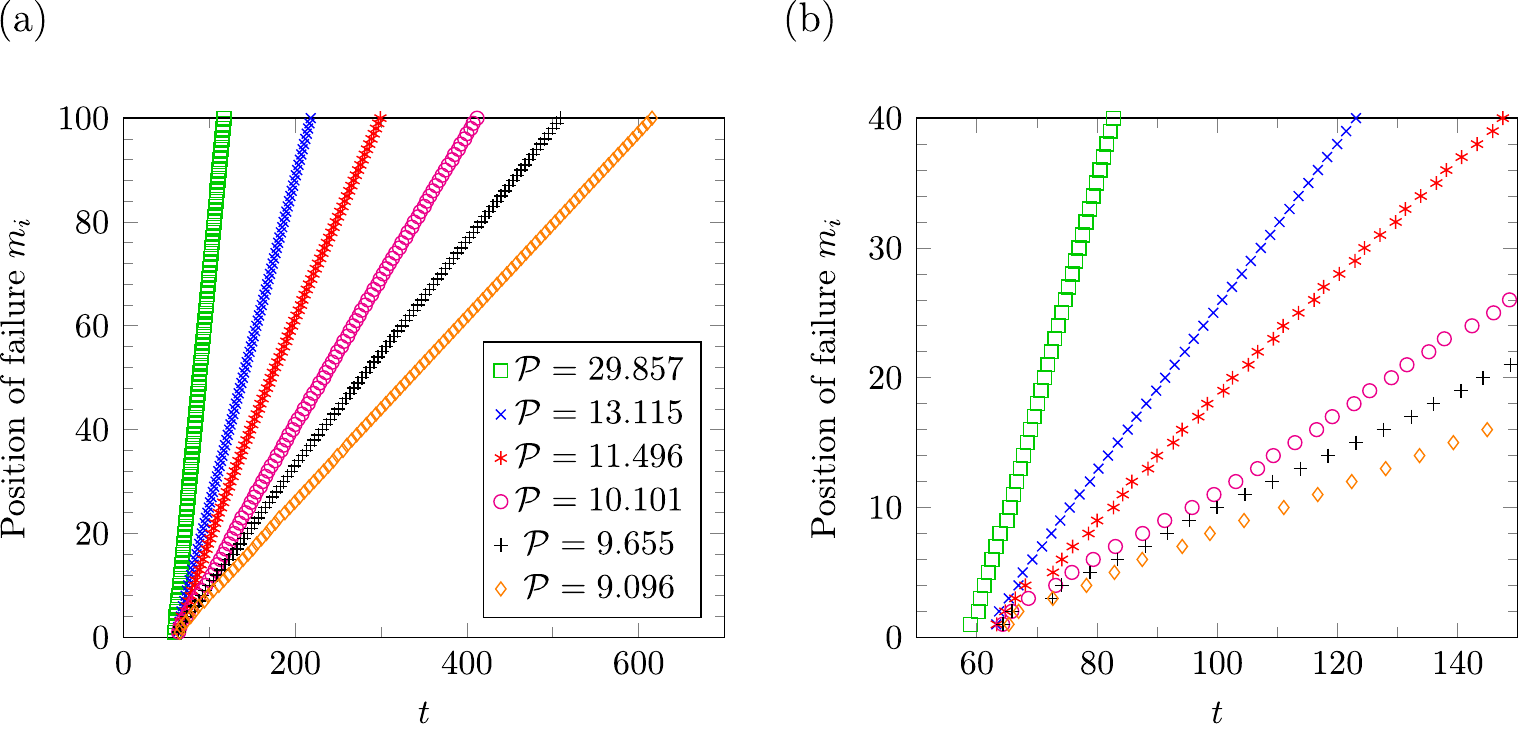}
\caption[]{
For $r=0.4$ and $\omega_0=3.1$, we show the behaviour of the transition front as a function of time for various load amplitudes $\CP$. (a) The position of failure inside the supported part of the structure as a function of time. (b) A magnification of the initial data of the computations  in (a). (online version in colour)
}
\label{Group1}
\end{figure}

 After these initial periods, the transition process settles and the position of failure appears to behave  as a linear function of time, representing that the steady-state failure phenomenon has been achieved. After the transient failure process, we can calculate the average failure speed from the data shown. The average speed~$\bar{v}$ is then computed as the slope of the line of best fit for this data. In this case, the average speeds obtained are $\bar{v}=0.1773, 0.2162, 0.2768, 0.3848, 0.6307, 1.7484$, that agree with the analytical predictions  for the possible failure speeds of  Section \ref{FW1.1}. As expected from physical considerations, Figure \ref{Group1} shows that as the load amplitude is increased, a higher failure speed is achieved.}

Actually, one can observe further oscillations in the failure speed when it appears to have settled to a uniform state.  We  define the instantaneous speed $v_i$ for the fracture process as
\begin{equation}
\label{eq3}
v_i=\frac{{m}_{i}-{m}_{i-1}}{{t}_{i}-{t}_{i-1}},  \text{ if } i\ge 2\;.
\end{equation}
 Using the notion of the instantaneous speed, we also can understand how quickly the failure process settles to the steady-state limit.

\begin{figure}[tp]
\centering
\includegraphics[width=14cm]{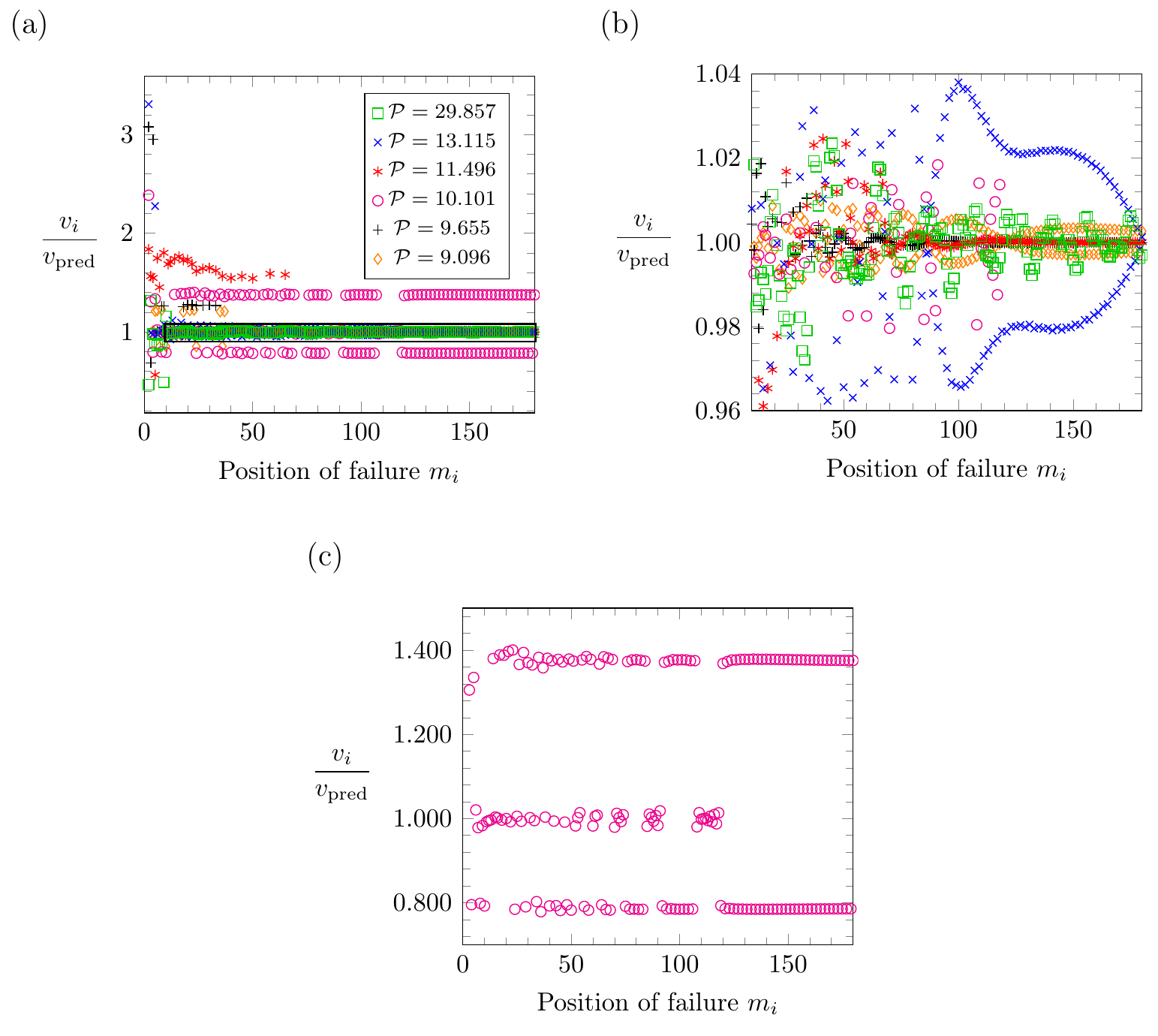}
\caption[]{ (a) The normalised instantaneous speeds computed via (\ref{eq3})  (and normalised by the respective predicted steady-state speeds based on the dispersion diagrams in Section \ref{secDISP}) as a function of the position of failure. (b) Magnification of the black box in (a). (c) A magnification of the normalised instantaneous speed distribution as a function of the position of failure for the case $\CP=10.101$.
The computations are performed for $r=0.4$ and $\omega_0=3.1$. (online version in colour) 
}
\label{Group1aa}
\end{figure}

In Figure~\ref{Group1aa}(a), we show the instantaneous speed plotted as function of the failure position $m_i$. Here, the instantaneous speed data presented corresponds to  cases considered in Figure \ref{Group1}. To demonstrate the speed of convergence to the steady-state failure process, each set of data for the instantaneous front speed has been normalised by the analytical prediction for the speed ${v_{\text{pred}}}$ that was attained in Figure \ref{Group1} (see Section \ref{FW1.1}).  In this case, convergence to the steady-state speed is represented by the convergence to unity in Figure \ref{Group1aa}.   
It is observed that the instantaneous speeds, for nearly all the load amplitudes, oscillate  about the predicted speed in each case (see the magnification in  Figure \ref{Group1aa}(b)). In fact,
Figure \ref{Group1aa}(b) shows that there exist micro-oscillations in speed of the failure front not necessarily seen in Figures \ref{Group1}(a) and (b). In addition, these oscillations in the transition front speed are not part of the analytical  model  studied in Section \ref{sec2}--\ref{sec5}, where the speed of the front is assumed to be uniform.

 Larger oscillations in the transition front speed are attained with larger values of the amplitude $\cal P$. The speed of convergence of these regimes appears to be a non-monotonic function of the load amplitude. 
  
  In Figure \ref{Group1aa}(c), we show a special case ($\CP=10.101$) where the instantaneous speed of the front does not converge to unity in the steady failure process, but instead settles to two distinct speeds (given by $v_i/v_{\rm pred}$ approximately equal to 0.79 and 1.39 after the 118$^{th}$ failure event). Although there is an apparent jump in the instantaneous speed of the failure, the average speed $\bar{v}$ is in fact equal to the predicted value 0.2768. Moreover, the corresponding data representing the failure position as a function of time in Figure \ref{Group1}(a)  exhibits a straight line that is commonly associated with the steady-state failure propagation.

\vspace{0.1in}\noindent {\bf \emph{Dependency of the average failure speed on the load amplitude} \rm} 

\vspace{0.1in}\noindent In the Figure~\ref{AmplitudeSpeed1}, 
 we show the average fracture speed~$\bar{v}$ as a function of the normalised oscillating force amplitude~{${\cal P}$}. 
There we see six plateaus corresponding to the steady-state speeds reached by the transition front observed in Figures \ref{Group1} and \ref{Group1aa}. These include the speeds~$\bar{v}=0.1773, 0.2768, 0.6307$, which correspond to alternating generalised strain regimes, and~$\bar{v}=0.2162, 0.3848, 1.7484$ that are the speeds associated with the pure steady-state regimes. 
 
 From Figure~\ref{AmplitudeSpeed1},  we can conclude:

 \begin{figure}[tp]
\centering
\includegraphics[width=8cm]{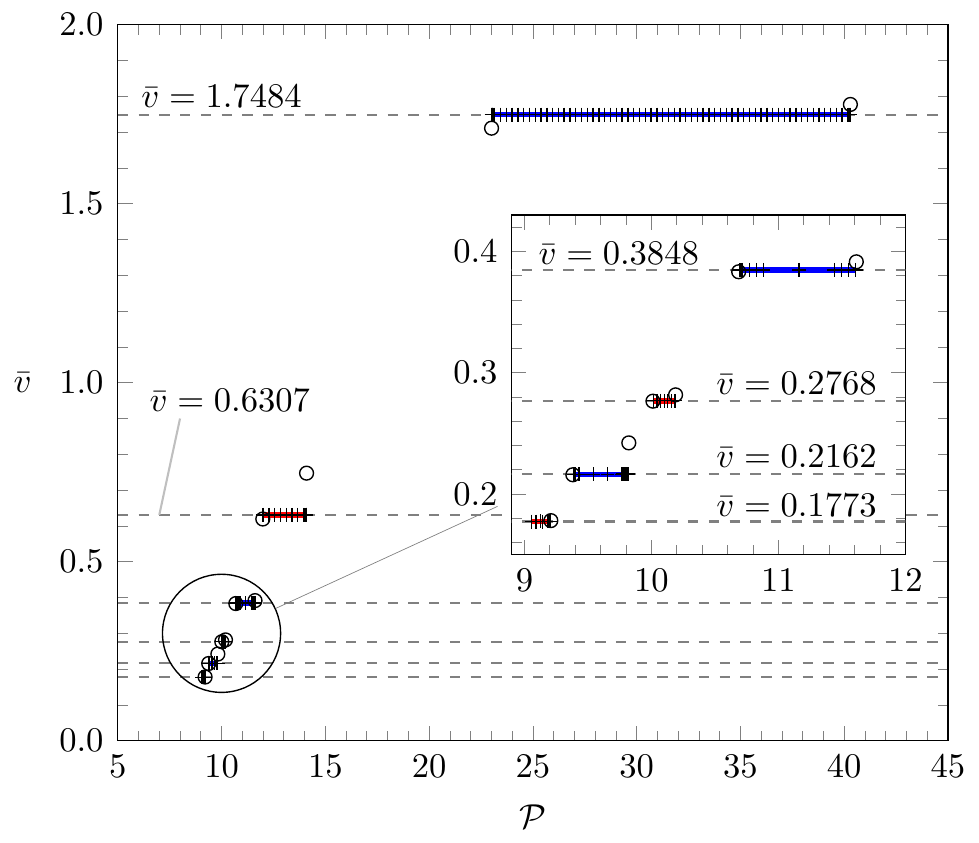}
\caption[]{The average failure speed shown as a function of the quantity ${\cal P}$, for $\omega_0=3.1$ and $r=0.4$. The circles and crosses represent the fracture behaviour for different regimes recorded from MATLAB simulations. Specifically, the crosses are associated with the steady-state propagation of the  transition front and the circles indicate  non steady-state propagation regimes. The dashed horizontal lines are for the predicted steady-state speeds that follow from  the  dispersion curve analysis in Section \ref{FW1.1}. Here, the plateaus for the steady-state speeds are shown as red for the alternating generalised strain regimes and  blue for the pure steady-state regimes. (online version in colour)
}
\label{AmplitudeSpeed1}
\end{figure}

\begin{enumerate}[$\bullet$]
\item The average speed is a monotonically increasing function of the load amplitude. 
\item There exists several plateaus, highlighted in red and blue indicating alternating generalised strain regimes and pure steady-state regimes,  respectively, where speeds predicted by the analytical model are realised. 
\item Only a finite collection of speeds predicted by the analytical model are realised and  lower speeds exist for narrow intervals of the load amplitude or are never realised.
\item Between any two pure steady-state regimes, corresponding to the fracture process associated with criteria (\ref{eq1}), there exists a plateau of finite width  at the failure speed  associated with the alternating generalised strain failure regime.
\item The size of plateaus for each steady-state regime corresponding to the alternating generalised strain fracture process increases with the load amplitude. The same behaviour is observed for the plateaus corresponding to the pure steady-state regimes.

\item Outside the plateaus, the fracture propagates non-uniformly. Between two steady-state speeds, we encounter the clustering phenomenon, first observed in \cite{MMS1}, where fracture propagates in regular periodic bursts. 
\end{enumerate}

\begin{table}
\centering
\begin{tabular}{c|c|c|ccc|}
\toprule
 \multicolumn{2}{c}{MATLAB results}
& {Analytical results} & \multirow{2}*{Regime}\\
{$\bar{v}$} 
&${\CP}_{\text{min}}$ & ${\CP}_{\text{min}}$& \\
\midrule
0.1773 & 8.985 &{8.932} & 1\\
0.2162 & 9.376  & 9.366  & 2\\
0.2768 & 10.045 &{10.055} & 1\\
0.3848 & 10.715 &10.686 & 2\\
0.6307 & 11.887  &{11.742}& 1\\
1.7484 & 23.048 &{23.00} & 2\\
\bottomrule
\end{tabular}
\caption{Load amplitude values that initiate steady-state failure regimes for $\omega_0=3.1$ and $r=0.4$.  The first two columns show the values for the average speed and the corresponding minimum load amplitude ${\CP_{\text{min}}}$ when the regime was first identified   from the numerical scheme of MATLAB. The third column presents the analytical predictions for the load amplitude based on the results of Section~\ref{sec4} and the theory of \cite{NMS, NMS2}. In the last column, we  specify the regime encountered, where ``1" represents the alternating generalised strain regime and ``2" indicates the pure steady-state regime.}
    \label{tab1}
\end{table}

{Additionally, with reference to Section \ref{sec5}, we show that the  analytical model can be used to efficiently predict the initial load amplitudes when steady-state failure regimes can appear.  In Table~\ref{tab1},  we present  data from the MATLAB  simulations when the  steady-state regimes are initiated and we supply the classification of each regime.  The analytical predictions for  these load amplitudes, {based on the right-hand side of (\ref{minloadpred})}  and the results of \cite{NMS, NMS2}, are also given.  We note  that there  is an excellent  match between the analytical results and those obtained from the numerical scheme implemented in MATLAB.} 

{Note that here we have shown only a finite collection of the predicted speeds is obtained in the simulations. The intervals of the load amplitude for when these steady-state speeds can occur have the form $[\CP_{\min}, \CP_{\rm max}]$. For the alternating generalised strain regime, $\CP_{\rm min}$ is determined using (\ref{minloadpred}) and (\ref{Pmin}), whereas for the pure steady-state regime it is determined using the results of \cite{NMS, NMS2}. On the other hand, the maximum value $\CP_{\rm max}$ of the load amplitude for which a steady-state regime can exist may be obtained by analysing the profile  of the structure based on the analytical solution, as shown in \cite{NMS2}. Here, for particular failure modes, as the failure speed decreases the distance between $\CP_{\rm min}$ and $\CP_{\rm max}$ becomes smaller  (for instance, see Figure \ref{AmplitudeSpeed1}).
In the case when the analytical model predicts $\CP_{\rm min}>\CP_{\rm max}$, which is a physically unacceptable scenario, the corresponding failure regimes cannot be realised.}

\vspace{0.1in}\noindent {\bf Behaviour of the system during the dynamic failure processes \rm}

\begin{figure}[tp]
\centering
\includegraphics[width=14cm]{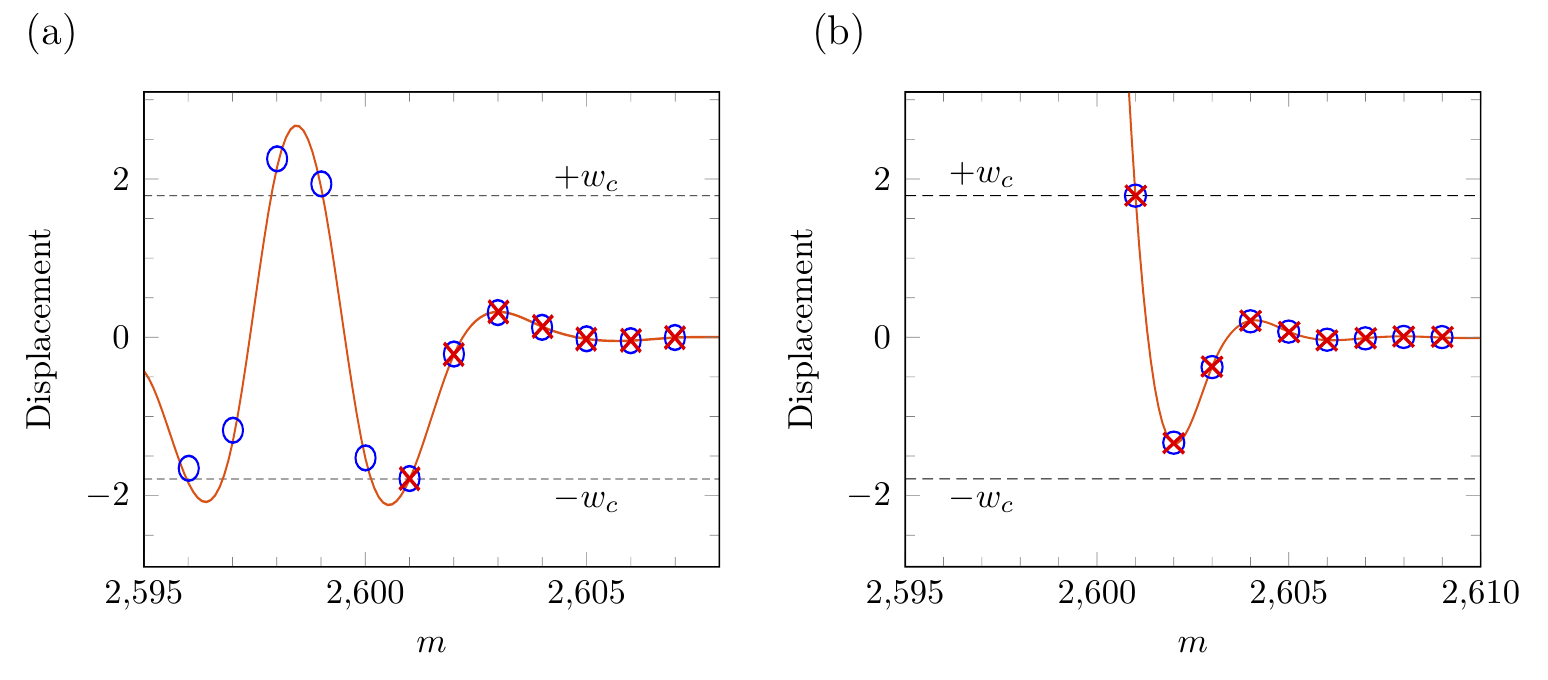}
\caption[]{Snapshots of the transition front steadily propagating, with two different regimes, through the structure excited by a harmonic load of frequency  $\omega_0=3.1$. In~(a) the snapshot is taken at $t={1012.4}$ for ${\cal P}=13.115$ and there the alternating generalised strain regime propagates with the average speed $\bar{v}=0.6307$. In~(b) $t={403}$ with ${\cal P}=33.484$ and this corresponds to a pure steady-state failure regime with the average speed $\bar{v}=1.7484$.
 The blue circles in  each panel are based on the   analytical results presented in Section \ref{sec4} and represent the mass displacements. The crosses indicate those masses situated in the supported region. The critical displacement for the links in the supported region is $w_c=1.7919$ and the limits corresponding to this value are shown by horizontal dashed lines.
The profiles shown are obtained from the MATLAB simulations by identifying the nodal displacements and rotations and using the results of Appendix A to reconstruct the deformation for the massless beams. In both computations the heterogeneity parameter $r=0.4$. (online version in colour)
 }
\label{Snapshot}
\end{figure}

\vspace{0.1in}\noindent For ${\cal P}=13.115$, Figure~\ref{Snapshot}(a) shows the system undergoing failure  at a particular time $t={1012.4}$ after the fracture of 600 pairs of transverse links.  The failure here propagates steadily under the alternating generalised strain regime. The average speed of this process observed from the MATLAB simulations is $\bar{v}=0.6307$.   In this case, one can observe the combination of the feeding and reflected waves  behind the transition front. Ahead of this point, there exists an evanescent wave. We note that at approximately $t=1034.0$, when the next    failure occurs, the profile of the structure can be obtained from that  shown in Figure~\ref{Snapshot}(a) mirrored about a horizontal line that corresponds to zero displacement.
In Figure~\ref{Snapshot}(b), we consider a pure steady-state regime resulting from when external load amplitude ${\cal P}=33.484$. The time when the snapshot is taken is  $t={403}$. The average steady-state speed achieved  is $\bar{v}=1.7490$ and as the failure propagates steadily, the profile of the structure  local to the transition front is preserved. Figure~\ref{Snapshot}(b) is an example of the pure steady-state regimes studied in~\cite{NMS, NMS2} that also occur in the simulations presented here.

In Figures~\ref{Snapshot}(a) and \ref{Snapshot}(b), we compare the MATLAB  simulation data for the arrangement of the masses along the profiles with the analytical results of Section~\ref{sec4} and \cite{NMS, NMS2}. In Figure \ref{Snapshot}(a), for the  computations based on the analytical results,  the inverse Fourier transform of (\ref{solp}) and (\ref{solm}) is computed using (\ref{C}), ensuring that (\ref{phidet}) holds for the feeding wave phase determined from (\ref{wc1}). For the computations in Figure \ref{Snapshot}(b), the same procedure is followed  based on the results of \cite{NMS, NMS2}.   We emphasise there is again an excellent agreement between the MATLAB results and the analytical predictions.

{\emph{ Failure induced by a high frequency.} \rm}\label{case1.2}
\begin{figure}[tp]
\centering
\includegraphics[width=14cm]{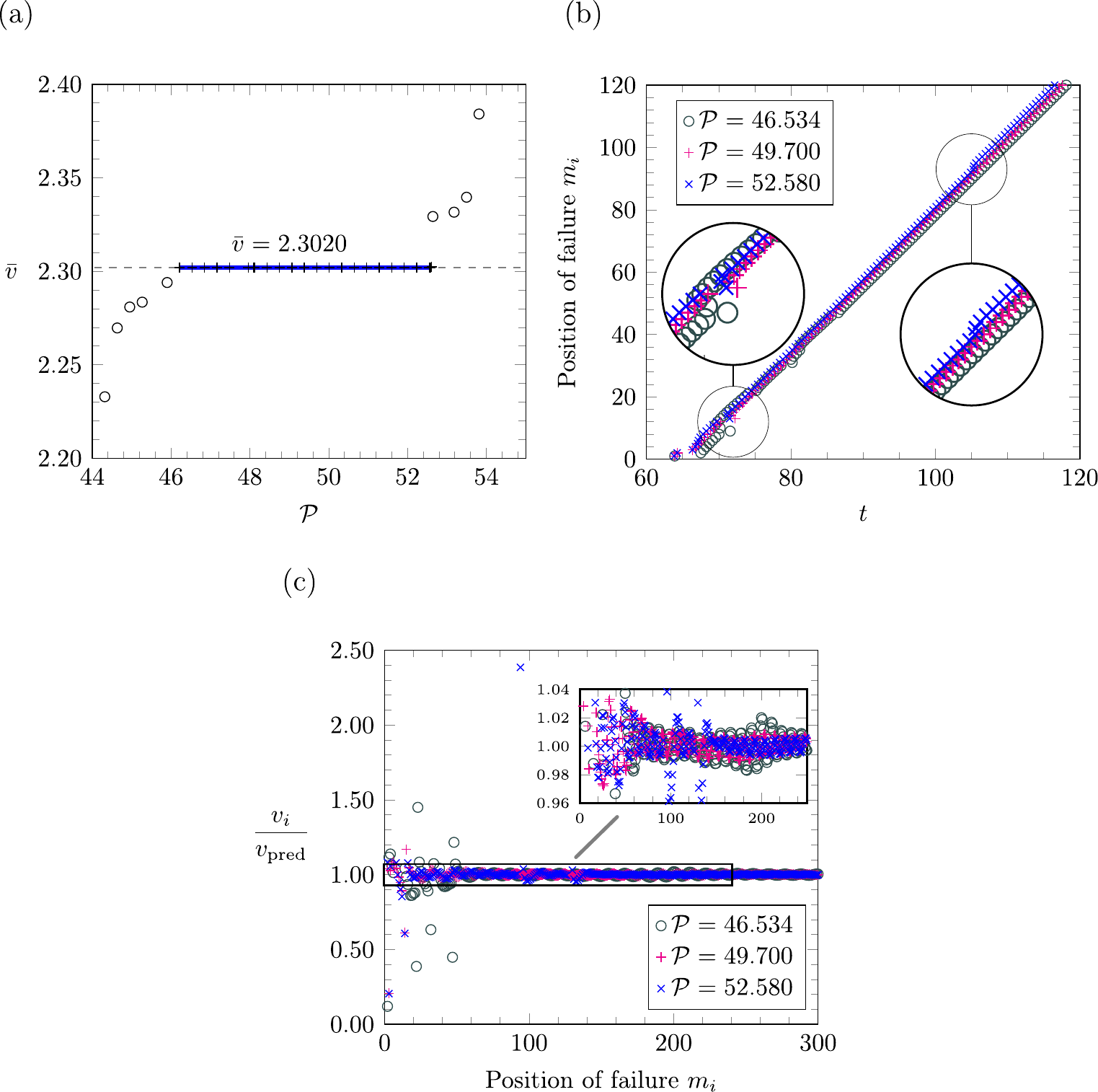}
\caption[]{Failure regimes identified for $\omega_0=5.9$ and $r=0.4$. (a) The average failure speed is shown as a function of ${\cal P}$. The crosses represent results of the MATLAB simulations where steady-state failure propagation is achieved, whereas as circles correspond to the non-steady failure regimes. The horizontal dashed lines are the predictions for the steady failure speeds based on the analytical model. In (b), we show the position of failure inside the supported part of the structure as a function of time. On the inset in (b), we present magnifications of the profiles where various non-steady propagation phenomena can be observed that are associated with the transient failure process. (c) The normalised instantaneous speed $v_i/v_{\text{pred}}$ computed using~\eqref{eq3} as a function of the failure position. (online version in colour)

}
\label{Group5}
\end{figure}
Here, we give an example of when the alternating generalised strain regime is never realised.
In this illustration, $\omega_0=5.9$ and $r=0.4$. As mentioned earlier in Section \ref{FW1.1}, the analytical model predicts a semi-infinite list of possible speeds for both the alternating generalised strain regimes and the pure steady-state regimes (also see the dispersion curves of Figure \ref{DispersionCurves1} with the intersection points for $\omega=5.9\, (=\omega_0)$). However, we observe only one of the predicted steady-state speeds in the MATLAB simulations. 

Figure~\ref{Group5}(a), shows the average speed  $\bar{v}$ as a function of the point load amplitude ${\cal P}$. This particular regime identified is a pure steady-state regime and the average failure speed is $\bar{v}=2.3020$. The load amplitude interval for where this regime is realised is $46.2172\le{\cal P}\le 52.612$. There is once again a good agreement with the  predictions of  analytical model in \cite{NMS, NMS2}.

In Figure~\ref{Group5}(b), we show the position of failure within the supported region as a function of the time  when ${\cal P}=46.534$, ${\cal P}=49.7$ and ${\cal P}=51.282$. These values are located within interval defining the plateau for the pure steady-state failure process. For ${\cal P}=46.534$, we see that the transition front speed is constant and equal to the predicted steady-state speed $v=2.3020$, after approximately  $44$ breakages. Prior to this,  the front propagates non-steadily in the transient regime. For ${\cal P}=49.7$, the pure steady-state regime is reached earlier at approximately the $13^{th}$ breakage. For ${\cal P}=51.282$, we see the failure inside the system settles to the steady-state regime after $100$ breakages.

For the same amplitudes, in Figure~\ref{Group5}(c), we present the instantaneous speed $v_i$, defined by~\eqref{eq3}, as a function of the position of fracture. 
The instantaneous speeds again oscillate about the speed predicted by the analytical model. It is apparent from  Figure~\ref{Group5}(c) that for those load amplitudes, situated at the extremes of the steady-state plateau in Figure \ref{Group5}(a), the failure process will behave irregularly during a large initial period of the failure process. For $\CP=49.700$, which is situated inside the plateau of Figure \ref{Group5}(a), the process is seen to converge to a more regular oscillatory behaviour earlier in the failure process.

\subsection{Case 2 - stiff supports}\label{FW2}

Here we analyse the failure process in a structure possessing transverse links  that are~``stiffer'' than the links along the central axis of the structure.

{\emph{Stiffer supports  and intermediate load  frequency.} \rm}
Here the supports in the system are characterised by the parameter $r=3.4$. We first  study  the case when this structure is subjected to an oscillating force with a  frequency~$\omega_0=~3.1$.
Although the flexural stiffness of the transverse links inside the supported region is now larger  compared to that of the beams along the central axis, the list of admissible steady-state speeds for failure, outlined in Section~\ref{FW1.1}, remains the same. The MATLAB simulations described below indicate which of these regimes are realised. 

\begin{figure}[tp]
\centering
\includegraphics[width=14cm]{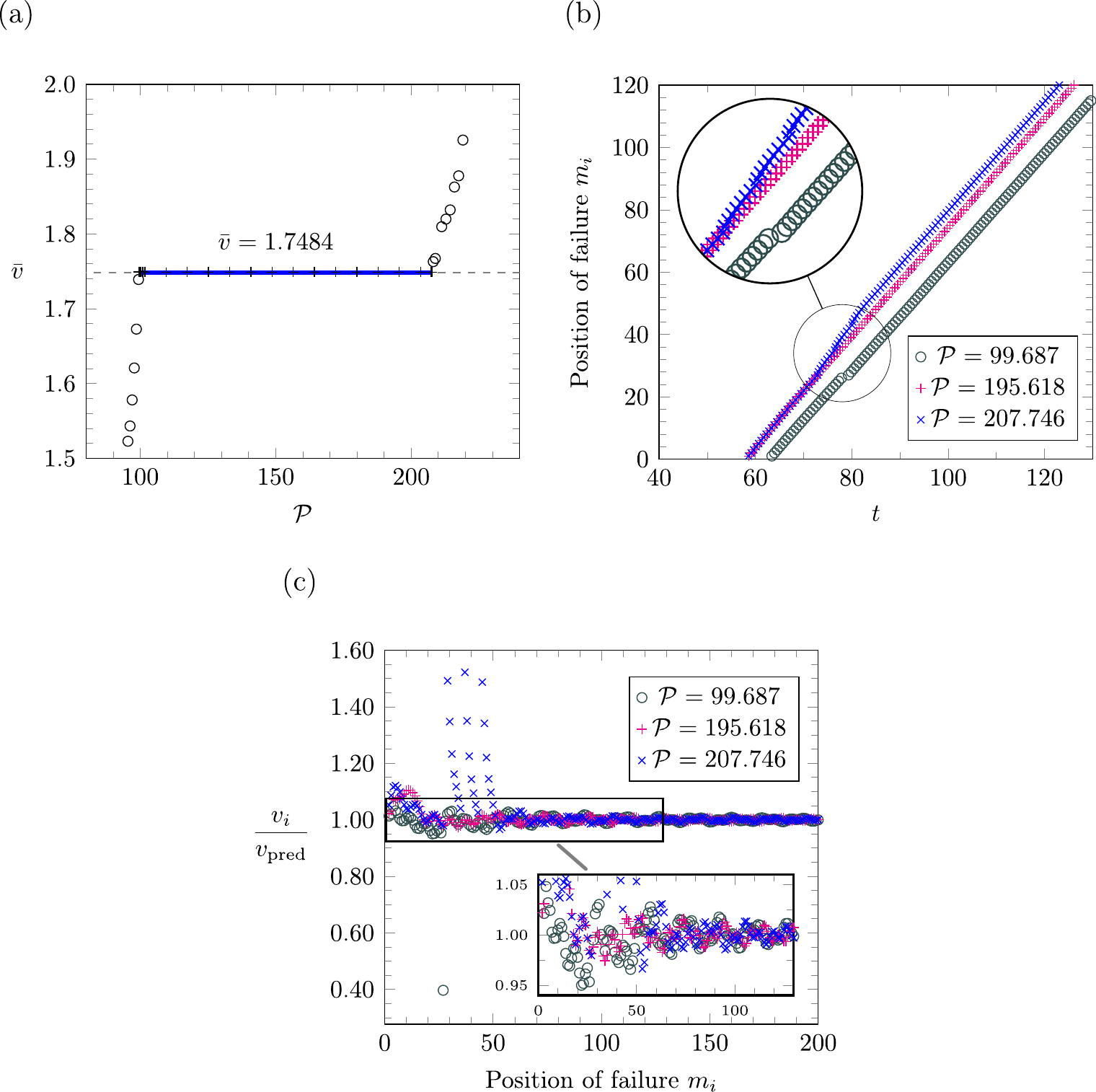}
\caption[]{Computations for the failure process in the structure with supports characterised by $r=3.4$ subjected to a load with frequency $\omega_0=3.1$. (a)  The average failure speed shown as a function of ${\cal P}$. Crosses represent the results of the MATLAB simulations where steady-state failure propagation is achieved. The circles correspond to the non-steady failure regimes. The horizontal dashed lines are the predictions of the failure speeds from the analytical model. In~(b), we show the position of failure inside the supported part of the structure as a function of time. (c) The normalised instantaneous speeds based on~\eqref{eq3} as a function of the position of failure. (online version in colour) 

}
\label{Group6}
\end{figure}

In Figure~\ref{Group6}(a), we show the average speed $\bar{v}$ as a function of $\cal P$. We recall the results of Section~\ref{subsec6}, where the same load frequency was considered and the structure was supported by transverse links with a lower flexural stiffness. There, it was possible to find six failure regimes (both alternating generalised strain regimes and pure steady-state regimes) with different speeds. 

When~$r=3.4$, we only observe one of the predicted steady-state speeds,~$v=1.7484$, which corresponds  to a pure steady-state regime with the highest possible speed. The regimes analysed in Sections~\ref{sec2}-\ref{sec4} are not observed.
In comparing with the computations of Figure \ref{AmplitudeSpeed1} for the case of a structure with softer supports, we note that  the stiffer supports require larger load amplitudes  to initiate and maintain the failure propagation, as expected.   Moreover, the stiffer supports allow for a larger range of load amplitudes where the highest steady-state speed can be observed (compare Figure \ref{Group6}(a) with Figure \ref{AmplitudeSpeed1}). The plateau in Figure \ref{Group6}(a) is defined by the interval {$99.452\le{\cal P}\le 211.972$}. This agrees with the predictions based on the results in \cite{NMS, NMS2}.

Figure~\ref{Group6}(b), shows the  position of failure against time for ${\cal P}=99.687, 195.62$ and $211.815$, located in the interval defining the plateau for the steady-state regime. For the load amplitude ${\cal P}=195.62$ the time interval for the initial transient failure process  appears to be small and the steady-state regime is realised with the average speed $\bar{v}=1.7484$. For ${\cal P}=99.687$, the transition front converges to the steady-state regime after approximately 27 breakages. For the largest amplitude ${\cal P}=211.815$, the steady-state failure process is achieved after approximately  48 breakages. It is interesting to note that for this larger load amplitude, the non-steady failure process attributed to the transient regime is visible and the profile propagates in steps that overlap (see {inset of Figure \ref{Group6}(b)}). This failure process is known as a forerunning fracture and was first identified in the study concerning the separation of a beam from an elastic foundation \cite{SASM}.

In addition, the normalised instantaneous speeds $v_i$, calculated using \eqref{eq3}, are shown in Figure~\ref{Group6}(c) as a function of the failure position. These computations reveal large fluctuations in the transition front speed about the analytically predicted speed  $1.7484$, (represented by unity on the vertical axis in this figure). It is evident that the failure process corresponding to the load amplitudes $\CP=99.687$ and $211.815$  (located at the extremes of the plateau in Figure \ref{Group6}(a))  give large fluctuations in the instantaneous speed of the transition front at the beginning of the failure process. After some time, 
for all cases shown,  the instantaneous speeds oscillate about the predicted speed value in a regular manner and the amplitude of the  oscillations decrease as the fracture process develops (see inset of Figure \ref{Group6}(c)).
 
 The example presented here shows that the number of failure regimes observed for different structures subjected to a remote sinusoidal load with a fixed frequency can vary. In this case, the stiffer supports have reduced the number of possible failure regimes that are realisable.

\begin{figure}[tp]
\centering
\includegraphics[width=14cm]{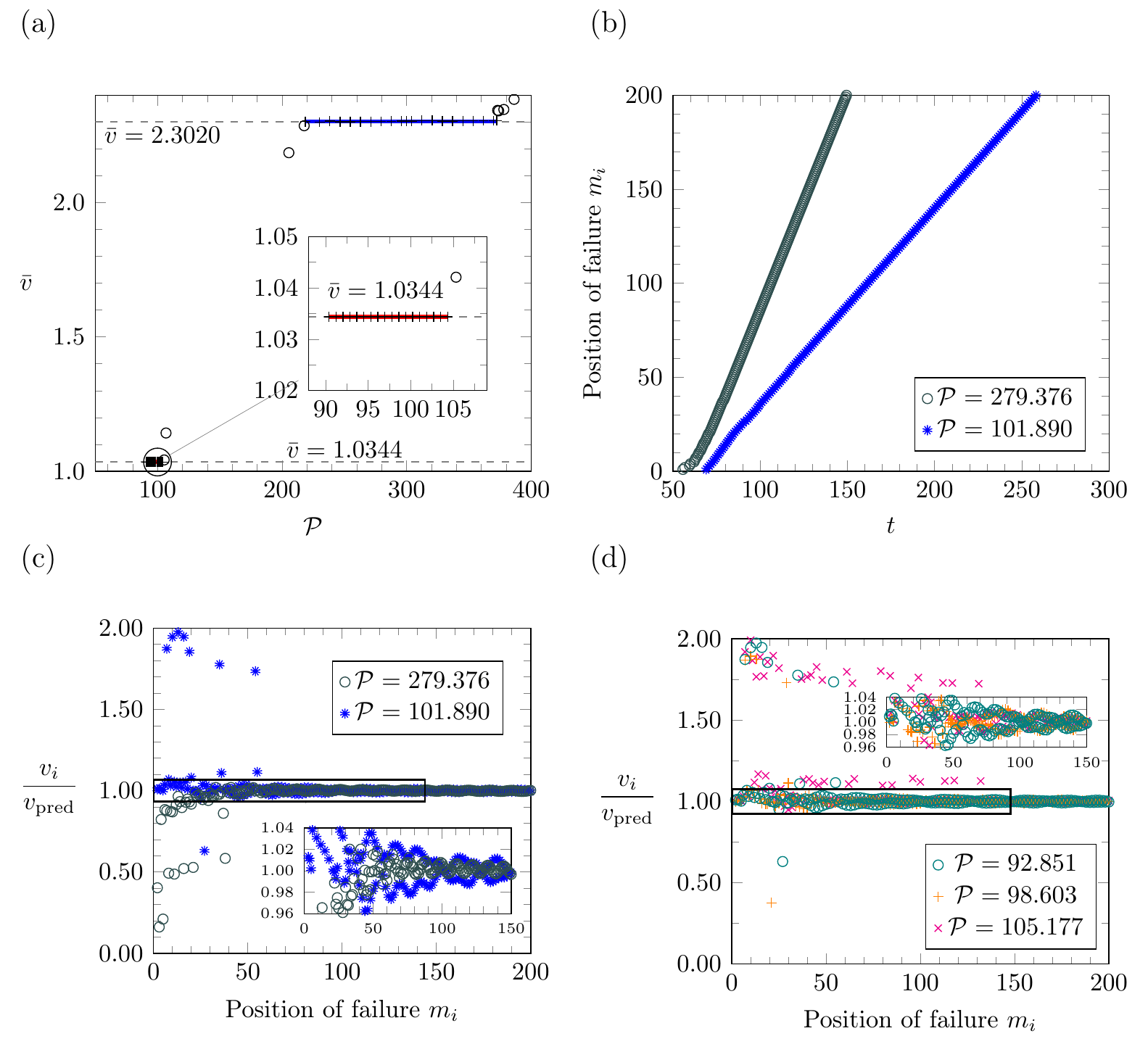}
\caption[]{Results showing the behaviour of the transition front in the MATLAB simulations for $r=3.4$ and $\omega_0=5.9$. In (a), the average failure speed shown as a function of ${\cal P}$. The crosses are the results of the MATLAB simulations where steady-state failure is achieved and the circles represent the non-steady failure regimes. The plateau indicated in red is for the alternating generalised strain regime, whereas the blue plateau is for the pure steady-state regime analysed in \cite{NMS, NMS2}. Horizontal dashed lines correspond to the predictions for possible failure speeds based on the analytical model.
In~(b), we show the position of failure as a function of time for load amplitudes situated inside each plateau  shown in (a). In (c), the  normalised  instantaneous speed distributions $v/v_{\text{pred}}$ are presented for the data in (b). In (d), for three load amplitudes in the plateau representing the alternating generalised strain regime we present the normalised instantaneous speeds of the failure. (online version in colour)}

\label{Group7}
\end{figure}

{\emph{High frequency loading of structure with stiff supports.} \rm}
Finally, we consider a structure with stiff supports and this structure is subjected to a sinusoidal load with a high frequency, which is located near the upper boundary of the passband for unsupported part of the structure.

This particular example shows that the alternating generalised strain regime can be realised in high-frequency loading configurations. In fact, the results of the  MATLAB simulations show there are two regimes encountered, as shown in Figure \ref{Group7}(a). In this figure, it can be seen that plateaus exist for $90.386\le{\cal P}\le105.341$ and ${218.57}\le{\cal P}\le376.664$. 
These intervals again agree well with the analytical predictions.
The plateau corresponding to the alternating generalised strain regime (with predicted speed $1.0344$) is very narrow in comparison to plateau belonging to the pure steady-state regime (having the predicted speed $2.3020$). If we compare with the computations of Figure \ref{Group1} for the same loading frequency, but for the structure with softer supports, we see the present configuration admits additional failure regimes at lower speeds. 


{Figure~\ref{Group7}(b) shows how the position of failure  varies as a function of time for load amplitudes chosen inside the plateaus of Figure \ref{Group7}(a). The non-steady behaviour of the transition front in the transient regime is noticeable and there exists a visible variation in the failure speed.  After this initial period, the system begins to settle to the steady-state failure process.}
The  instantaneous speed distribution obtained using~\eqref{eq3} for the cases analysed in Figure \ref{Group7}(b) is presented in Figure~\ref{Group7}(c).
It is again evident that there are large oscillations in failure speed about the analytically predicted values in the initial period of the failure process. When the failure process settles,  the instantaneous speeds follow a regular oscillatory pattern about the predicted speed. In this example, the computations for the pure steady-state regime ($\CP=279.376$) appear converge the fastest to the predicted speed (see inset of Figure \ref{Group7}(c)) than those for the alternating generalised strain regime.  
Finally, we comment on the instantaneous speed distributions, shown in Figure \ref{Group7}(d),  for some load amplitudes within the lowest plateau corresponding to the alternating generalised strain regime. 
The instantaneous speeds associated with $\CP =92.851$ and $\CP=105.177$, at the extremes of the load amplitude interval defining this plateau, clearly require a large duration of time to settle to the steady-state failure speed. In fact in the case of $\CP=105.177$, several higher speeds are observed up to the 140$^{th}$ instant of failure. Following this, the instantaneous speed eventually settles into steady oscillations about the predicted speed from the analytical model for all cases shown in Figure \ref{Group7}(d).

\section{Conclusions}\label{Conclusions}
We have developed an analytical model characterising a particular propagation regime in a flexural system, composed of beams connecting periodically placed masses. The regime identified corresponds to the scenario where the generalised strains, represented by bending moments and shear forces in the beam connections, alternate in sign as the failure advances through the system.

Analytical results characterising the dynamics of the system during its failure and  when possible failure regimes occur have been determined. In particular, the results have been shown to provide  excellent predictions for the behaviour of sufficiently long finite systems subjected to sinusoidal loads that were modelled with a numerical scheme developed in MATLAB.
These simulations  revealed that the existence of regimes modelled here is dependent on the parameters describing the structure and sinusoidal loading. In addition, the  failure regimes identified in \cite{NMS, NMS2} can also be realised  together with those studied here in the MATLAB simulations.

The model can be generalised to consider different types of loading and more complicated failure criteria, where other types of special failure phenomena may exist. Applications of the tools developed here are envisaged in civil engineering, where the dynamics and failure of large multi-structures  are considered.

\vspace{0.1in}{\bf Acknowledgements: \rm}M.J.N. gratefully acknowledges the support of the EU H2020 grant MSCA-IF-2016-747334-CAT-FFLAP. 
I.S.J.  would like to thank the EPSRC (UK) for its support through Programme Grant no. EP/L024926/1.


\setcounter{equation}{0} \renewcommand{\theequation}{A.\arabic{equation}}
\section*{Appendix}
\renewcommand{\thesubsection}{\Alph{subsection}}
\subsection{Derivation of the governing equations in terms of generalised coordinates of the masses}
The displacements $W_m$ and $W^{\pm}_j$,  along the horizontal and transverse beams, respectively,  can be found by solving the equations
\[\frac{{\rm d}^4W_m}{{\rm d}\tilde{x}^4}(\tilde{x}, t)=0\;, \quad 0<\tilde{x}<a \;, m\in \mathbb{Z}\]
and
\[\frac{{\rm d}^4W^{\pm}_j}{{\rm d}\tilde{y}_{\pm}^4}(\tilde{y}_{\pm}, t)=0\;, \quad 0<\tilde{y}_\pm<a \;,  j>\lfloor Vt/a\rfloor \;, j \in \mathbb{Z}\;,\]
for $t>0$, $\tilde{x}=x-am$, $\tilde{y}_+=y$ and $\tilde{y}_-=y+a$. The  function $W_m$ satisfies the boundary conditions
\[W_m(0, t)=w_m(t)\;, \quad  \frac{{\rm d}W_m}{{\rm d} \tilde{x}}(0, t)=-\theta_m^y(t)\;, \quad W_m(a, t)=w_{m+1}(t)\;, \quad \frac{{\rm d}W_m}{{\rm d} \tilde{x}}(a, t)=-\theta_{m+1}^y(t)\;.\]
The function $W_j^{+}$ is subject to the conditions
\[W_m^{+}(0, t)=w_m(t)\;,  \quad \frac{{\rm d}W_m^{+}}{{\rm d} \tilde{y}}(0, t)=\theta_m^x(t)\;, \quad W^{+}_m(a, t)=0\;,\quad \frac{{\rm d}W^{+}_m}{{\rm d} \tilde{y}}(a, t)=0\;, \]
whereas  $W_j^{-}$ satisfies
\[W_m^{-}(0, t)=0\;, \quad  \frac{{\rm d}W_m^{-}}{{\rm d} \tilde{y}}(0, t)=0\;, \quad W^{-}_m(a, t)=w_m(t)\;, \quad \frac{{\rm d}W^{-}_m}{{\rm d} \tilde{y}}(a, t)=\theta_m^x(t)\;. \]
Thus, from the above we have
\begin{eqnarray}
\!\!\!\!\!\!\!\!\!\!\!\!W_m(\tilde{x}, t)&=&-{[a(\theta_m^y(t)+\theta_{m+1}^y(t))-2(w_{m}(t)-w_{m+1}(t))]}\frac{\tilde{x}^3}{a^3}\nonumber \\
&&+[a(2\theta_m^y(t)+\theta_{m+1}^y(t))-3(w_m(t)-w_{m+1}(t))]\frac{\tilde{x}^2}{a^2}-\theta^y_m(t)\tilde{x}+w_m(t)
\label{Wmform}
\end{eqnarray}
and
\begin{eqnarray}
W_m^{+}(\tilde{y}_+, t)&=&[a \theta_m^x(t)+2w_m(t)]\frac{\tilde{y}_+^3}{a^3}-[2a\theta^x_m(t)+3w_m(t)]\frac{\tilde{y}_+^2}{a^2}+\theta^x_m(t) \tilde{y}_++w_m(t) \;,\nonumber\\
\label{Wp}
\\
W_m^{-}(\tilde{y}_-, t)&=&[a \theta_m^x(t)-2w_m(t)]\frac{\tilde{y}^3_-}{a^3}-[a\theta_m^x(t)-3 w_m(t)]\frac{\tilde{y}^2_-}{a^2}\;.\label{Wm}
\end{eqnarray}
{Note, in the considered phase transition problem, the transverse supports fail at the connections to rigid interface. In this case, behind the transition front, the transverse beams are attached to the masses at one end and are free at the other. At the free end, the bending moments and shear forces are zero. Hence, the displacements along these beams can be shown to be linear functions of their local $y$-co-ordinate. Consequently, they do not contribute to the linear and angular momentum balance of the masses behind the transition front (for instance, see (\ref{eqa})).}

The bending moments in the transverse beams applied about the $x$-axis can be calculated via
\[ \mathcal{M}^{x,\pm}_m(\tilde{y}_{\pm}, t)=E_2I_2\frac{{\rm d}^2W^{\pm}_m}{{\rm d}\tilde{y}_{\pm}^2}(\tilde{y}_{\pm}, t)\;.\]
The angular momentum balance about the $y$-axis for a mass in the supported region  then gives
\[ \mathcal{M}^{x,+}_m(0, t)- \mathcal{M}^{x,-}_m(a, t)=0\;,\]
where the right-hand side is zero as it is assumed all masses have negligible moments of inertia. 
This together with (\ref{Wp}) and (\ref{Wm}) implies $\theta^x_m=0$ in the supported region.
The shear forces in these beams  then have the form
\[\CV_m^{y, \pm}(\tilde{y}_{\pm}, t)= -E_2I_2 \frac{{\rm d}^3W^{\pm}_m}{{\rm d}\tilde{y}_{\pm}^3}(\tilde{y}_{\pm}, t)\]
and using  (\ref{Wp}) and (\ref{Wm}) leads to (\ref{TF}). 
The internal bending moments and shear forces in the $m^{th}$ beam aligned with the horizontal axis, (see Figure \ref{Figura1}) are computed via
\begin{equation}
\CM_m^y(\tilde{x}, t)=-E_1I_1\frac{{\rm d}^2W_m}{{\rm d} \tilde{x}^2}(\tilde{x}, t) \quad \text{ and }\quad  \CV_m^y(\tilde{x}, t)=-E_1I_1\frac{{\rm d}^3W_m}{{\rm d} \tilde{x}^3}(\tilde{x}, t)\;, 
\end{equation}
respectively. Insertion of (\ref{Wmform}) into these relations yields (\ref{eq1My}) and (\ref{eq1Vy}).

\setcounter{equation}{0} \renewcommand{\theequation}{B.\arabic{equation}}
\subsection{An oscillating point force in an infinite  mass-beam chain}
Here, we derive the amplitude associated with the outgoing waves generated by point force in a mass-beam chain. The force is assumed to have frequency $\omega_0$ and amplitude $P_0$. According to the derivation of the governing equations in Section \ref{sec2.2}, the problem may be written as
\begin{eqnarray}\label{eq1geA}
\frac{Ma^3}{E_1 I_1}\f{\D^2{w}_m(t)}{\D t^2}&=&-{6}\left\{2[2{w}_m(t) -{w}_{m-1}(t)-{w}_{m+1}(t)]-a[{\theta}^y_{m+1}(t)-{\theta}^y_{m-1}(t)]\right\}\nonumber \\  &&+\frac{a^3}{E_1 I_1}P_0 \sin( \omega_0 t)\delta_{m, 0}
\end{eqnarray}
and
\begin{equation}\label{eq2geA}
3[{w}_{m+1}(t)-{w}_{m-1}(t)]+a[{\theta}^y_{m+1}(t)+{\theta}^y_{m-1}(t)+4{\theta}^y_m(t)]=0\,,
\end{equation}
Here the Kronecker delta $\delta_{i, j}$ has been used to represent the position of the force in the structure.
We consider the complex solution of this problem, i.e. we look for the solutions as
\begin{equation}\label{complex_wth}
w_m(t)=\text{Re}(\mathfrak{w}_m(t))\;, \quad \theta^y_m(t)=\text{Re}({\vartheta}^y_m(t))
\end{equation}
which allows us to consider the problem in terms of complex functions $\mathfrak{w}_m$ and $\vartheta^y_m$:
\begin{eqnarray}\label{eq1geAA}
\frac{Ma^3}{E_1 I_1}\f{\D^2{\mathfrak{w}}_m(t)}{\D t^2}&=&-{6}\left\{2[2{\mathfrak{w}}_m(t) -{\mathfrak{w}}_{m-1}(t)-{\mathfrak{w}}_{m+1}(t)]-a[{\vartheta}^y_{m+1}(t)-{\vartheta}^y_{m-1}(t)]\right\}\nonumber \\  &&-\frac{a^3}{E_1 I_1}{\rm i}P_0 e^{\I \omega_0 t}\delta_{m, 0}
\end{eqnarray}
and
\begin{equation}\label{eq2geAA}
3[{\mathfrak{w}}_{m+1}(t)-{\mathfrak{w}}_{m-1}(t)]+a[{\vartheta}^y_{m+1}(t)+{\vartheta}^y_{m-1}(t)+4{\vartheta}^y_m(t)]=0\,.
\end{equation}
Here we non-dimensionalise the equations by introducing
\[\omega_0=\sqrt{\frac{E_1I_1}{Ma^3}}\tilde{{\omega}}_0\;, \quad P_0=\sqrt{\frac{E_1I_1}{a^2}}\tilde{P}_0\;, \quad   \mathfrak{w}_m=a \tilde{\mathfrak{w}}_m\;,\]
and we will also assume the solutions take the form
\begin{equation}\label{timeharmonic}
\tilde{\mathfrak{w}}_m(t)=(-1)^m \CW_m e^{{\rm i} \omega_0 t}, \qquad {\theta}^y_m(t)=(-1)^m\Theta^y_m e^{{\rm i} \omega_0 t}\;.
\end{equation}
We drop the ``tilde" in going forward and  derive the system
\begin{eqnarray}\label{eq1geA}
-\omega_0^2\mathcal{W}_m(t)&=&-{6}\left\{2[2{\CW}_m +{\CW}_{m-1}+{\CW}_{m+1}]+[{\Theta}^y_{m+1}-{\Theta}^y_{m-1}]\right\}-{\rm i}P_0\delta_{m, 0}
\end{eqnarray}
and
\begin{equation}\label{eq2geA}
-3[{\CW}_{m+1}-{\CW}_{m-1}]+4{\Theta}^y_m-{\Theta}^y_{m+1}-{\Theta}^y_{m-1}=0\,,
\end{equation}
where all variables in the above equations are dimensionless.
In the following, the  discrete Fourier transforms are used
\[{\CW}^F=\sum_{m=-\infty}^\infty \CW_m e^{{\rm i}k m}\; \quad \text{ and } \quad {\Theta}^{y, F}=\sum_{m=-\infty}^\infty \Theta^y_m e^{{\rm i}k m}\;.\]
We take the discrete Fourier transform to obtain
\begin{eqnarray}\label{eq1geA1}
[-24(1 +\cos(k))+\omega_0^2]{\CW}^F+12{\rm i}\sin(k){\Theta}^{y, F}-{\rm i}P_0=0\,,
\end{eqnarray}
and 
\begin{equation}\label{eq2geA1}
6\text{i}\sin(k){\CW}^F+2(2-\cos(k)){\Theta}^{y, F}=0\,.
\end{equation}
Combining (\ref{eq1geA1}) and (\ref{eq2geA1}) we have
\[\CW^F=-\frac{{\rm i}P_0}{h_2(k, {\text{i}\omega_0})}\;\]
and applying the inverse discrete Fourier transform yields
\[\CW_m=-\frac{{\rm i}}{2\pi}\int_{-\pi}^\pi\frac{P_0e^{-\text{i} k m}}{h_2(k, {\text{i}\omega_0})}dk\;.\]
For outgoing waves from the source to exist we require that $\omega_0<\sqrt{48}$, representing  the pass band of the structure. In this case, in the interval $-\pi<k<\pi$ the chosen load frequency can be linked with two wave numbers $\pm k_f\mp \text{i}0$, that define the waves propagating in the medium.  These wavenumbers are simple poles of the kernel in the above integral. The above is then rewritten  as
\[
\CW_m=-\frac{{\rm i}}{2\pi}\int_{-\pi}^\pi\frac{P_0e^{-\text{i} k m}}{(0-\text{i}(k-k_f))(0+\text{i}(k+k_f))R_0(k, {\omega_0})}dk\;, \]
where the positive function $R_0$ is defined in (\ref{R0def}) and has no zeros on the real axis. 
Applying the residue theorem,  under the assumption $m \to \infty$, the integral on the right-hand side has the asymptote
\[\frac{1}{2\pi}\int_{-\pi}^\pi\frac{P_0e^{-\text{i} k m}}{(0-\text{i}(k-k_f))(0+\text{i}(k+k_f))R_0(k, {\omega_0})}dk\sim- \frac{{\rm i}P_0 e^{-\text{i}k_f m}}{2 k_f R_0(k_f, \omega_0)}\;,\quad \text{ for }m\to \infty\;.\]
Finally, combining  this with (\ref{timeharmonic}) and (\ref{complex_wth}) leads to  
\[w_m(t)\sim  \frac{P_0}{2k_f R_0(k_f, \omega_0)}\cos((k_f+\pi)m-\omega_0 t-\pi) \quad \text{ for }\quad  m\to \infty\;.\]
Thus, from this we see that the amplitude of the feeding wave in the problem considered in Section \ref{sec5} is
\[A=\frac{P_0}{2k_fR_0(k_f, \omega_0)}\;,\]
where the variables in this  expression are dimensionless.
\end{document}